\documentclass[10pt,journal,compsoc]{IEEEtran}

\usepackage{xspace}
\usepackage{color}
\usepackage{tabularx}
\usepackage{listings}
\usepackage{graphicx}
\usepackage{tabularray}
\usepackage{booktabs}
\usepackage{amssymb}
\usepackage{tikz}
\usetikzlibrary{positioning, arrows.meta, backgrounds, calc, fit}            
\usepackage{tcolorbox} 
\tcbuselibrary{theorems}

\usepackage{algorithm}
\usepackage[noend]{algpseudocode}

\ifCLASSOPTIONcompsoc
 \usepackage[caption=false,font=normalsize,labelfont=sf,textfont=sf]{subfig}
\else
 \usepackage[caption=false,font=footnotesize]{subfig}
\fi

\usepackage{url}

\begin{document}
%
\title{\FT: Agentic Hardware-in-the-Loop Fuzzing for 5G Devices \nsurevise{and beyond}}

	

%
\IEEEoverridecommandlockouts
\author{\IEEEauthorblockN{
        Guoqiang Hua\IEEEauthorrefmark{1}\IEEEauthorrefmark{2},
        Sakshi Udeshi\IEEEauthorrefmark{2},
        Matheus E. Garbelini\IEEEauthorrefmark{2},
        Daisuke Mashima\IEEEauthorrefmark{1} and
        Sudipta Chattopadhyay\IEEEauthorrefmark{3}}\\
        \IEEEauthorblockA{\IEEEauthorrefmark{1}Singapore University of Technology and Design (SUTD)}
        \IEEEauthorblockA{\IEEEauthorrefmark{2}Temasek Laboratories, Nanyang Technological University}
        \IEEEauthorblockA{\IEEEauthorrefmark{3}University of Missouri-Kansas City}\\
        \IEEEauthorblockA{Email: guoqiang\_hua@mymail.sutd.edu.sg\IEEEauthorrefmark{4}\thanks{\IEEEauthorrefmark{4}Corresponding author.},
    \{sakshi.udeshi, matheus.garbelini\}@ntu.edu.sg, \\
    daisuke\_mashima@sutd.edu.sg,
    sudiptac@ieee.org}
}



\newcommand{\FT}{\textsc{5uGue}\xspace}
\newcommand{\FG}{\textsc{5Ghoul}\xspace}
\newcommand{\todo}[1]{\textcolor{red}{\textbf{TODO: }\emph{#1}}}

\newcommand{\var}[1]{\textit{#1}}                 
\newcommand{\proc}[1]{\textsc{#1}}                
\newcommand{\tool}[1]{\texttt{#1}}                
\newcommand{\kw}[1]{\textbf{#1}} 

\newcommand{\marker}[1]{\texttt{#1}}


\newcommand{\revise}[1]{\textcolor{black}{#1}}
\newcommand{\screvise}[1]{\textcolor{black}{#1}}
\newcommand{\grevise}[1]{\textcolor{black}{#1}}
\newcommand{\mrevise}[1]{\textcolor{black}{#1}}


\newcommand{\nsurevise}[1]{\textcolor{black}{#1}}
\newcommand{\ngrevise}[1]{\textcolor{black}{#1}}
\newcommand{\scnrevise}[1]{\textcolor{black}{#1}}
\newcommand{\mnrevise}[1]{\textcolor{black}{#1}}
\newcommand{\scnnrevise}[1]{\textcolor{black}{#1}}
\newcommand{\nngrevise}[1]{\textcolor{black}{#1}}


\newcommand{\tsegrevise}[1]{\textcolor{black}{#1}}
\newcommand{\screv}[1]{\textcolor{black}{#1}}

\newcommand{\claude}{Claude Sonnet 4.6\xspace}
\newcommand{\gpt}{GPT-5.5\xspace}
\newcommand{\gemini}{Gemini 3.1 Pro\xspace}
\newcommand{\ds}{DeepSeek V4 Pro\xspace}
\newcommand{\kimi}{Kimi K2.6\xspace}
\newcommand{\qwen}{Qwen3.7 Plus\xspace}

\newcommand{\circled}[1]{\textcircled{\raisebox{-0.9pt}{#1}}}

\algnewcommand{\LineComment}[1]{\State \(\triangleright\) \textcolor{purple}{#1}}

\maketitle


\begin{abstract}



In this paper, we present \FT, a large language model (LLM)-driven hardware-in-the-loop agent, which enables the wireless fuzzing of 5G-enabled devices (such as mobile phones). \FT combines multi-step reasoning with tool-assisted interaction and
integrates into the fuzzing pipeline by leveraging specialized tools for internet search, 5G protocol analysis, memory management, and test-case execution. 
The agent iteratively generates test cases, executes them on targets, 
collects runtime feedback, and adaptively refines the fuzzing strategy.
We evaluate \FT on five commercial off-the-shelf (COTS) 5G devices across six LLMs. 
The best performing runs for \FT trigger 167 crashes in total (141 \ngrevise{novel} and 31 unique) within 60 interaction turns (49 to 133 minutes total runtime).  
In contrast, the best runs for the state-of-the-art
baseline (\FG) discovers only 13 crashes in total (10 \ngrevise{novel} and 6 unique) within 150 minutes. 
\scnrevise{Overall, by integrating LLM-driven agents, \FT 
demonstrates significantly better crash discovery performance 
(12.8× total, 5.2× unique improvement).} \tsegrevise{Notably, our experiments identified 46 unique crashes, 
44 of which were confirmed by the vendors but had been previously reported. 
More importantly, MediaTek confirmed two \emph{previously unknown, high-severity vulnerabilities}, 
one of which was assigned \texttt{CVE-2026-20525}.}
\nsurevise{Finally, to demonstrate that \FT is agnostic to 
wireless protocols, we also extend it to 
Bluetooth Classic with only minimal code changes. In our 
evaluations, 
\ngrevise{\FT finds an average of 6.4 novel 
crashes and 1.6 unique crashes.}}
%
In a nutshell, \FT highlights that LLM agents are capable to operate effectively beyond software-in-the-loop settings, extending autonomous vulnerability discovery to the proprietary, hardware-bound systems that affect billions of end-user devices \nsurevise{across multiple wireless protocols}.
%
%
%
\end{abstract}



%

\section{Introduction}


The security of 5G-enabled devices is a problem of enormous scale.
By the end of 2025, there \grevise{were} an estimated 2.9 billion 5G subscriptions,
accounting for one-third of all mobile subscriptions, and this is projected
to reach 6.4 billion by 2031, or two-thirds of all mobile subscriptions~\cite{ericsson-5g-forecast}. 
Each of these devices relies on proprietary technologies processing untrusted
over-the-air signaling, so a single class of vulnerability can impact a vast,
rapidly growing population of users. Yet the black-box, hardware-bound nature
of these devices makes them difficult to analyze at scale, leaving this
critical attack surface comparatively \screvise{underexplored} in the context of offensive security. 

Large Language Models (LLMs) have demonstrated strong capabilities in
code generation, program analysis, and vulnerability discovery.
While agentic systems enhance these capabilities through autonomous task execution, existing LLM-based agents are predominantly designed for software-in-the-loop settings~\cite{mayoral2025cai, simsek2025pocgen}. 
These software-in-the-loop agents often have source code access, coverage and feedback to steer the fuzzing.
This is in contrast to real-world wireless systems (such as the 5G baseband devices), 
where analysis mostly needs to be performed in black-box settings on proprietary hardware.
As a result, this \emph{rapidly growing} attack surface remains out of reach for state-of-the-art fuzzing agents and applying LLM-based agents to \emph{wireless 5G security} analysis remains an open, active and emerging space of work.


\begin{figure}[t]
    \centering
    \includegraphics[width=\linewidth]{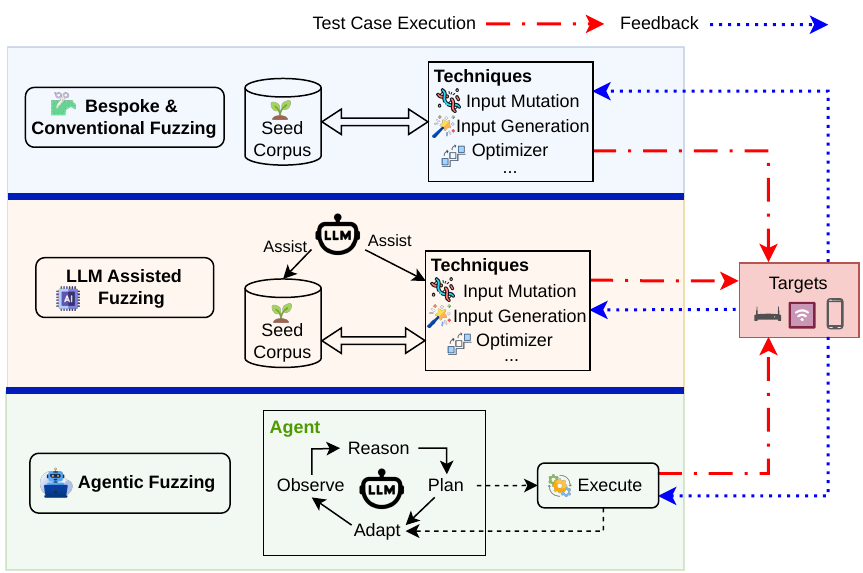}
    \caption{\FT with respect to state-of-the-art approaches}
    \label{fig:intro}
\end{figure}


\smallskip \noindent
To address this gap, we propose \FT \footnote{Pronounced ``fugue''; a play on 5G and User Equipment (UE), and a nod to the fugue state the \FT agent induces in 5G devices}, an autonomous \emph{hardware-in-the-loop} agent that fuzzes commercial-off-the-shelf (COTS) 5G devices over the air (OTA). \nsurevise{Though we have developed \FT primarily for the 5G protocol, the agentic loop 
makes no protocol-specific assumptions. 
To showcase this, we extend \FT to Bluetooth Classic (BT) by making minimal procedural code changes (see \ngrevise{Section~\ref{sec:bt_result})}}.
We view the state of fuzzing through three paradigms (Figure~\ref{fig:intro}): 
\begin{itemize}
    \item \textbf{Bespoke and conventional fuzzing} is fully directed by the creators of the system. Usually, a custom engine~\cite{5ghoul,llfuzz,U-fuzz,5G-aka-bypass} 
mutates an initial seed corpus or \screvise{manipulates communication on-the-fly} with certain heuristics. Coverage and crash feedback are then looped back into the mutator. \screvise{While such systems have shown promising outcomes in the form of discovering new vulnerabilities, they are 
    \emph{pre-encoded} by human experts.}
    \item \textbf{LLM-assisted fuzzing} inserts an LLM as a \emph{component}, not a strategist. The LLM is typically leveraged to strengthen a single constituent element of the fuzzing pipeline 
    e.g., to generate seeds~\cite{chatafl}, mutate inputs as per a specification~\cite{llmif} \screvise{or} guide the existing fuzzing setup~\cite{attention_distance}. The model sharpens parts of the fuzzing workflow but it does not \emph{see, or steer} the 
    entire fuzzing loop. 
    \item In contrast, \FT casts the LLM as a \emph{strategist} with visibility over the full \screvise{fuzzing} loop. The agent \nsurevise{sources the initial seeds, plans tests, executes these tests on a COTS 5G or BT device} over the air, observes the resulting logs and possibly crash signals, reasons over it, and determines what to try next, all while having the ability to adapt the strategy based on data through modular tools (See Figure~\ref{fig:intro}: \textbf{Agentic Fuzzing}). 
    The \FT agent orchestrates the full fuzzing loop without any human intervention. 
\end{itemize}

This work makes the following contributions:
\begin{itemize}
    \item \textbf{The first agentic, hardware-in-the-loop fuzzing framework for COTS 5G \grevise{devices}:} We design and implement \FT. To the best of our knowledge, this is the first agentic fuzzing framework that operates directly on COTS 5G devices. We create an extensible and modular architecture. All the capabilities (e.g., exploit execution, memory, web/literature search) are self contained tools that can be added and removed with minimal effort, making \FT easy to extend and ablate. 
    \item \textbf{Evaluations:} We evaluate \FT along multiple axes. Specifically,
    \begin{enumerate}
        \item \textbf{Tooling:} We evaluate three progressively improved \FT versions and show that adding specific tools enhances fuzzer performance \screvise{in discovering modem crashes.}
        \item \textbf{Models:} \screvise{We demonstrate \FT is 
        LLM-agnostic by using six state-of-the-art LLMs.}
        \item \textbf{Generality:} We fuzz five COTS 5G devices showcasing \FT's generality and real world performance.
        \item \textbf{Protocol Extensibility:} \nsurevise{We extend and evaluate \FT beyond 5G to Bluetooth Classic (BT) by making only minimal code changes (i.e. modifying the exploit tool and prompts).}
    \end{enumerate}

    \FT demonstrates significantly better crash discovery performance. Specifically the best \FT runs discover \textbf{12.8×} the total crashes over the best state-of-the-art baseline runs (see Section~\ref{sec:eval-results}). 
    
    \item \textbf{Confirmed Vulnerabilities:} 
    \nsurevise{We have evaluated phones across two modem vendors, and both vendors have confirmed our findings.}
    \tsegrevise{Our experiments identified 46 \textbf{unique} crashes: 38 in MediaTek modems and 8 in UNISOC modem. 
    Among these, 44 crashes (36 from MediaTek and 8 from UNISOC) were confirmed by the vendors, 
    although they had previously been reported by other researchers. 
    More importantly, MediaTek confirmed two \textbf{previously unknown, high-severity vulnerabilities} 
    discovered in our experiments, one of which was assigned \texttt{CVE-2026-20525}.}
    \tsegrevise{Notably, the identified vulnerabilities stem from genuine modem-level flaws 
    and can be exploited to trigger denial-of-service (DoS) attacks.}
    
    \item \textbf{Interactive Dashboard:}  We release an interactive dashboard cataloging every test run for our evaluations, to aid the transparency and reproducibility of our framework. This dashboard can be found at:
    \begin{center}
        \ngrevise{\textbf{\url{https://5ugue-evaluations.pages.dev}}}
    \end{center}

\end{itemize}

\section{Background}
\smallskip\noindent
\textbf{Cellular System}: In this paper, we focus on 5G New Radio (NR) cellular systems. 
A \textit{base station} provides radio access between the telecommunication network 
and 5G devices such as smartphones. In 5G NR, this base station is referred to as a \textit{gNB}, 
which implements the radio access functions required for communication with user equipment over the NR air interface. 
Moreover, \textit{user equipment} (UE) refers to an end-user device that connects to a cellular network. 
In 5G NR, communication is organized into multiple protocol layers~\cite{3gpp_ts_38300}, including the physical layer, 
which handles radio transmission, the Media Access Control (MAC) layer, which manages resource scheduling, 
and the Radio Link Control (RLC) layer, which performs packet segmentation, 
along with other layers such as PDCP and RRC.
Our experiments mainly focus on the non-physical layers, from the MAC layer upward, 
particularly on \textbf{pre-authentication} 5G messages exchanged before the security context is established. 

\smallskip\noindent
\textbf{Threat Model:}
We assume that the attacker sets up a rogue gNB, i.e., she can 
simulate a 5G NR base station and connect to a victim UE over the air. The attacker is modeled as a Dolev-Yao adversary: 
it can \mrevise{eavesdrop}, modify and inject network messages but it cannot decrypt encrypted messages without the key.
Prior works~\cite{5ghoul,llfuzz,5greasoner} have shown that 
such an attack model \screvise{is practical to launch   
pre-authentication attacks}. 
\screvise{Similar to several prior works~\cite{5ghoul,llfuzz}, 
we leverage a software defined radio (SDR) to simulate 
a rogue gNB.}

\begin{figure}[t]
    \centering
    \includegraphics[width=\linewidth]{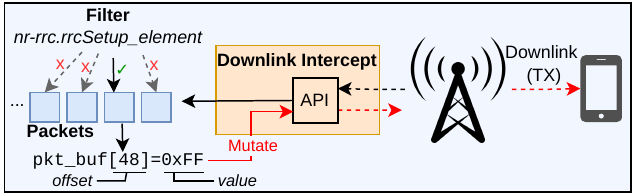}
    \caption{Example Test Case}
    \label{fig:5g_test_case}
\end{figure}

\begin{figure}[t]
    \centering
    \includegraphics[width=\linewidth]{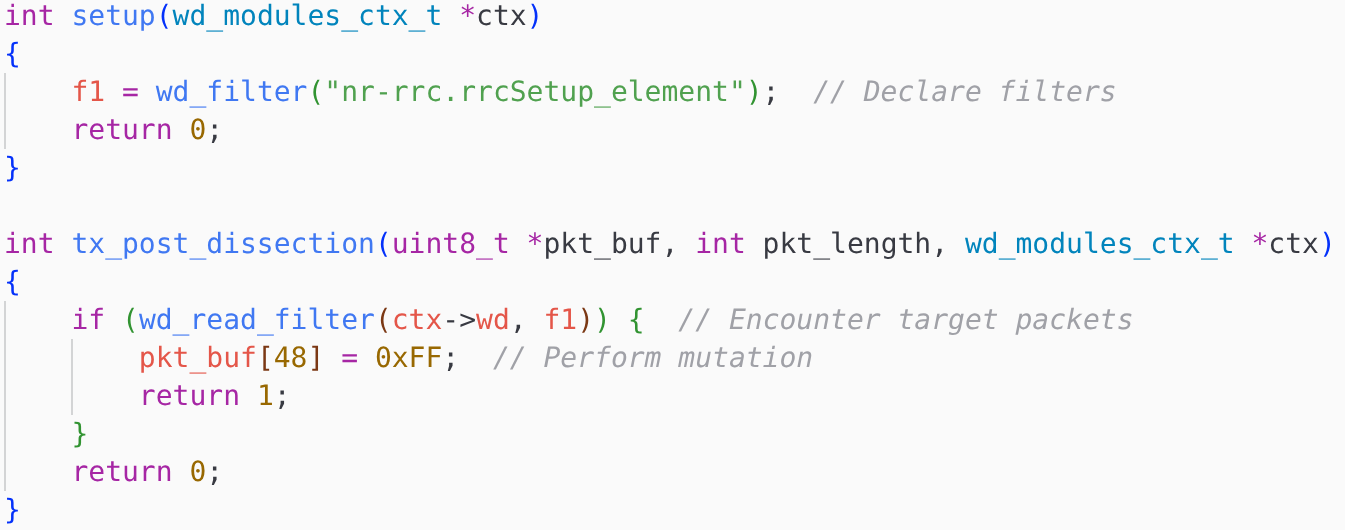}
    \caption{Example Test Case Code}
    \label{fig:5g_test_case_code}
\end{figure}

\smallskip\noindent
\textbf{Anatomy of a 5G NR test case:}
\screv{We leverage tools employed in prior works~\cite{5ghoul,U-fuzz} 
to act as a man-in-the-middle (MitM) and mutate the payload of selected packets. 
As illustrated in Figure~\ref{fig:5g_test_case}, a test case consists of 
a Wireshark display filter and one or more \textit{offset-value} mutations. 
The filter selects the downlink packets (\screv{i.e., packets sent from a 5G base station to a 5G user equipment}) to intercept, 
while each \textit{offset-value} pair specifies where and how the packet payload is modified.
In this example, \FT targets packets containing an RRCSetup element using the 
filter \texttt{nr-rrc.rrcSetup\_element} and replaces the 48\textsuperscript{th} byte ({\em offset}) with \texttt{0xFF} ({\em value}). 
The interception interface applies the mutation before transmitting the modified packet to the target device.
Figure~\ref{fig:5g_test_case_code} shows the corresponding C++ test case generated by \FT. 
The \texttt{setup} function declares the packet filter, while
\texttt{tx\_post\_dissection} checks each outgoing packet against the filter and applies
\texttt{pkt\_buf[48] = 0xFF} when a match is found. 
\FT then compiles and executes the generated test case and 
examines the resulting device logs and runtime feedback for potential crashes.}

\smallskip\noindent
\textbf{LLM-based Agentic Systems}:
%
An agentic LLM system couples an LLM with the ability to take actions, and gather feedback from an external environment, 
in contrast to only emitting text in response to a prompt~\cite{rise_of_agents_survey, llm_agents_survey}. 
An agentic framework allows the model to be treated as a decision making controller in the 
\emph{Observe}~$\rightarrow$~\emph{Reason}~$\rightarrow$~\emph{Plan}~$\rightarrow$~\emph{Execute}~$\rightarrow$~\emph{Adapt} 
loop (See Figure~\ref{fig:intro}).

In modern LLM-driven agentic systems, this capability is realized through tool use, where the model invokes external functions such as search engines, code compilers, or domain-specific APIs and incorporates these results into the system's 
decision making~\cite{react_agents, anthropicBuildingEffective}. To improve over repeated attempts, agents additionally employ various 
techniques. These include self-reflection where the agents analyze failures and feed the results 
back into later decisions~\cite{reflexion}. An agentic system can carry out
multi-step tasks with limited human supervision, which makes it well suited to problems
that require iterative exploration and adaptation. 
In this work, we explore the potential of such agentic systems for hardware-in-the-loop fuzzing of COTS 5G devices. For a broader overview, we refer interested readers to the following resources on agentic systems~\cite{rise_of_agents_survey, llm_agents_survey, anthropicBuildingEffective, anthropic_context_engineering, openaiHarnessEngineering, agents_software_engineering}.
\section{Related Work}



\smallskip\noindent
\textbf{Conventional Software Fuzzing:}
Vulnerabilities in software are critical and can be exploited by attackers 
en-masse to create harm. Vulnerabilities like WannaCry ransomware affected about 230,000 to 300,000 computers~\cite{wannacry_crisis}, causing an estimated massive financial impact of USD 4 Billion~\cite{wannacry_financial}. Comparable incidents, such as the Heartbleed flaw in OpenSSL, left an estimated 500,000 publicly trusted server exposed~\cite{heartbleed_impact}.
%
Fuzzing, which was introduced 
in the early 1990s~\cite{fuzzing_intro_reliability}, 
remains of the most widely deployed techniques to assess software 
reliability. 
As of June 2026, GitHub alone hosts about 11,000 repositories related to fuzzing~\cite{github_fuzzing_repositories_2026}. 
For a detailed treatment of classic software fuzzing we refer the readers to the following surveys~\cite{art_science_of_fuzzing, survey_software_fuzzing, demystify_fuzzing}. In the following, 
we provide a brief overview of the emerging field of LLM/Agentic Software Fuzzing and the state-of-the-art for over-the-air 
5G fuzzing. 

\smallskip\noindent
\textbf{Agentic Software Fuzzing:}
%
One of the early efforts in LLM-powered security testing is \textit{PentestGPT}~\cite{pentestgpt}, which adopts a feedback-driven, modular architecture.
Next, \textit{ChatAFL}~\cite{chatafl} uses LLMs to extract protocol information, diversify seed
messages, propose next message / sequences to explore new protocol states, thus achieving higher code coverage than {\em AFLNet}~\cite{aflnet}.
\textit{LLMIF}~\cite{llmif} extends LLM-driven fuzzing to wireless targets (i.e., Zigbee) by feeding constraints context, examples, and execution feedback to the LLM so that produced tests remain compliant to the protocol while generating error-inducing inputs.
\textit{mGPTFuzz}~\cite{mGPTFuzz}
leverages LLMs to parse 1000+ pages of Matter protocol specification into machine-usable
guidance and then performs stateful sequence generation.
Finally, \textit{ELFuzz}~\cite{elfuzz} automatically synthesizes
generation-based fuzzers tailored to a system under test via LLM driven synthesis over
a fuzzer space. 
We also refer interested readers to the these 
relevant surveys~\cite{fuzzing_meets_llms, llm_driven_fuzzing}.

In contrast to the aforementioned works, \FT performs \emph{hardware-in-the-loop} \emph{over-the-air (OTA)} fuzzing 
of COTS 5G NR devices. Every test exercises real silicon 5G modems over a software-defined-radio test-bed.
The \FT agent acts as the principal strategist in a closed loop. \FT plans each test, 
delivers the mutated message over a radio link to a physical handset, and re-plans 
using the vendor-reported crash oracle. 
{\em To the best of our knowledge, \FT is the first agentic, hardware-in-the-loop fuzzer to operate
directly against commercial 5G baseband firmware.} 


\smallskip\noindent
\textbf{5G NR Testing and Fuzzing:}
Several works adapted software fuzzing to the 5G wireless domain, all of which focus on different aspects of mobile protocols. Firstly, LTEFuzz~\cite{ltefuzz} introduced negative testing towards 4G Base Station (eNB) via generation of invalid RRC and NAS uplink payloads, discovering many implementation bugs on commercial eNBs. Next, \FG~\cite{5ghoul} was the first work to enable directed fuzzing of layer 2 5G protocols such as MAC, RLC and PDCP. 
It serves as the basis for \FT due to its \screvise{comprehensive documentation and maintenance}~\cite{5ghoul-github}. Other works such as LLFuzz expanded fuzzing to lower layers of 4G protocols by mutating less known signaling messages towards the UE. 
Lastly, CONSET~\cite{conset} \screvise{generates test cases} by extracting 
specification-level constraints that violate semantics of 
ASN.1 fields 
\mrevise{
through LLM-inferred DSL 
rules.
However, such work is tailored to only mutate messages that contain ASN.1 fields, which are not present in lower layer 5G protocols (i.e., MAC, RLC, PDCP). In contrast, \FT generates invalid messages across any 5G protocol layer regardless of the presence of ASN.1 fields and it also extends 
easily to other protocols (e.g., BT).
\screv{Moreover, we view CONSET
as a complementary work that can be integrated in a future 
extension of the open platform provided by \FT.}}


Compared to these works, \FT provides a full end-to-end, agentic OTA fuzzing design that automatically produces and verify attack payloads crashing 5G COTS devices. \FT also serves as a platform that can be applied to other wireless technologies by simply changing the testbed and agent tools.

\smallskip\noindent
\textbf{Hardware Fuzzing:}
\screvise{Our work is orthogonal to the line of work that aims 
to fuzz hardware designs written in certain hardware description 
language~\cite{genhuzz,hwfuzz-2022-usenix}. In contrast to this line of 
work, \FT focuses on an end-to-end OTA fuzzing with 5G COTS devices 
in the loop.}


\section{Methodology}

In this section, we elucidate the design and implementation of \FT. The goal of \FT is to investigate if we can design an LLM-based 
\emph{agentic system} to enhance over-the-air (OTA) fuzzing. 
We enable the \FT to act as a strategist equipped with a set 
of tools such as {\em web search}, {\em persistent memory}, 
{\em over-the-air delivery} and {\em exploit compilation} 
leveraging an LLM. This enables the discovery of novel crashes 
in closed-source 5G New Radio (NR) modems. \FT is a closed-loop, 
feedback driven fuzzer. It is not used just as a mutation 
generator; instead, \FT acts as the principal strategist
by selecting 
\screvise{concrete test cases (packet filters, offsets and values)} 
and re-planning after each test using structured feedback 
(i.e., crash identifier, mutation count, packet count, and 
state-machine diff).
\nsurevise{Additionally, unlike a traditional fuzzing framework 
where the \scnrevise{seeding strategy} is usually fixed, 
\FT treats the strategy for seeding its \emph{initial fuzzing 
inputs} as an alterable component.}

In the subsequent sections, we 
delve into the details of \FT design. 


%




\subsection{System Architecture}

\begin{algorithm}[t]
\caption{\FT (V3) main loop. One iteration of the outer \kw{while} is a \emph{turn}.}
\label{alg:fuzzagent-turn}
\begin{algorithmic}[1]
\Require strategy memo \var{S}, version config \var{C}, capture path \var{P}, modem profile \var{M}
\Ensure  per-turn updates to \var{msgs}, \var{kb}, and crash record
\Statex \emph{Let} $\mathcal{T} = \{\tool{execute\_test},\,\tool{execute\_test\_batch}\}$
\State \var{turn} $\gets 0$;\quad \var{firstCrash} $\gets$ \kw{false};\quad \var{fails} $\gets 0$
\While{\kw{true}} \label{alg:fuzzagent-turn:start-while}
    \State \var{turn} $\gets$ \var{turn} $+\,1$
    \If{\var{maxTurns} $> 0$ \kw{and} \var{turn} $>$ \var{maxTurns}}
        \State \kw{break} 
    \EndIf \label{alg:fuzzagent-turn:reach-max-turn}
    \State \var{S} $\gets$ \var{kb}.\proc{GetStrategy}()
    \State \var{prompt} $\gets$ \proc{BuildPrompt}(\var{S},~\var{C},~\var{P},~\var{firstCrash},~\var{M}) \label{alg:fuzzagent-turn:build-prompt}
    \State \proc{SummarizeIfNeeded}(\var{msgs}) \label{alg:fuzzagent-turn:summarize}
    \State \var{resp} $\gets$ \proc{LLM.Chat}(\var{prompt},~\var{msgs},~\var{toolSchemas}) \label{alg:fuzzagent-turn:llm-call}
    \State \var{msgs}.\proc{Append}(\var{resp}.\var{assistant})
    \ForAll{\var{tc} $\in$ \var{resp}.\var{toolCalls}} \label{alg:fuzzagent-turn:tool-call-start}
        \State \var{out} $\gets$ \proc{Dispatch}(\var{tc}.\var{name}, \var{tc}.\var{args}) \Comment{run tool}
        \State \var{msgs}.\proc{Append}(\proc{ToolResult}(\var{tc}.\var{id},~\var{out})) \label{alg:fuzzagent-turn:tool-call-res}
        \If{\var{tc}.\var{name} $\in \mathcal{T}$} \label{alg:fuzzagent-turn:tool-call-test}
            \State \var{kb}.\proc{LogTest}(\proc{Parse}(\var{out}))
            \LineComment{consecutive tests with no crash/new state}
            \State \proc{UpdateFailures}(\var{fails},~\var{out}) 
            \State \proc{MaybeMarkFirstCrash}(\var{firstCrash},~\var{out})
        \EndIf
    \EndFor
    \If{\var{firstCrash} just confirmed}
        \LineComment{redirect to novel crashes}
        \State \var{msgs}.\proc{Append}(\proc{PostFirstCrashDirective}) \label{alg:fuzzagent-turn:post-first-crash}
    \EndIf
    \If{\var{fails} $\geq 5$}
        \LineComment{reflect in the event of consecutive failures}
        \State \var{msgs}.\proc{Append}(\proc{ReflectionPrompt}) 
        \State \var{fails} $\gets 0$ \label{alg:fuzzagent-turn:fail-to-0}
    \EndIf
\EndWhile
\end{algorithmic}
\end{algorithm}

\FT's main loop is organized as follows: 
\begin{enumerate}
 \itemsep0em 
    \item \FT rebuilds a system prompt with the current strategy memo, version configuration, packet capture path, and a post-first-crash flag 
    (\grevise{Line~\ref{alg:fuzzagent-turn:build-prompt} in  Algorithm~\ref{alg:fuzzagent-turn}}). 
    The post-first-crash flag indicates whether the fuzzer has found a crash or not. 
    \item Check conversation length and summarize if needed, \grevise{as indicated in Line~\ref{alg:fuzzagent-turn:summarize}}. 
    \item Invokes a single \texttt{chat.completions.create} call (using an OpenAI-compatible completions API) that carries the full tool schema list and the running message history, \grevise{as demonstrated in Line~\ref{alg:fuzzagent-turn:llm-call}}.
    \item Once the LLM API call is completed, for each tool call, dispatch to the appropriate tool handler 
    \grevise{as shown in Line~\ref{alg:fuzzagent-turn:tool-call-start} to Line~\ref{alg:fuzzagent-turn:tool-call-res}}. 
    \item In case there exists a request to \emph{execute} one or multiple tests (run an exploit), \FT commits the test results to a knowledge base. This may also trigger an update for the unique crash set, self reflection (on five consecutive failures) or update the previously mentioned post-first-crash flag.
    Specifically, 
        a ``failure'' occurs when the test compiles, but 
        neither triggers a crash nor observes a new state in the 
        fuzzing. 
        Once the post-first-crash flag is true, we inject a \emph{POST-FIRST-CRASH} directive to instruct the agent to focus on novel cases, as a result of the crash pipeline being validated. \grevise{This is shown in Line~\ref{alg:fuzzagent-turn:tool-call-test} to Line~\ref{alg:fuzzagent-turn:fail-to-0}.}
    \item \FT continues this loop till the maximum defined turns (see \grevise{Lines~\ref{alg:fuzzagent-turn:start-while}-\ref{alg:fuzzagent-turn:reach-max-turn} in Algorithm~\ref{alg:fuzzagent-turn}}).
\end{enumerate}


\smallskip\noindent
\textbf{Context management:}
The \FT system estimates the token count by converting message chars to tokens at a 4:1 ratio~\cite{openai_tokens}. When the estimate exceeds a pre-set, \revise{version dependent threshold (see Table~\ref{tab:agent-versions})}, the oldest message up to the last six messages are condensed into a summary. The summarizer module just does not truncate the messages, it walks the tool calls and tool results, parses for the `\marker{test id}', `\marker{crash triggered}' to emit `\marker{Test Results}' and `\marker{Key Findings Recorded}', 
\screvise{respectively}. The \FT summarizer uses `\marker{finding id}' markers to emit a section named `\marker{Agent Reasoning Highlights}'.  

\smallskip\noindent
\textbf{Strategy memory:}
LLM agents today do not have a reliable memory layer~\cite{memGPT}. 
As a result, \FT does not rely only on conversation history. Instead, 
a plain-text markdown ``\marker{strategy}'' memo is persisted across turns. This strategy memo 
is re-injected at the end of every system prompt. This lets the agent maintain a durable plan e.g., a sweep tracker of tested 
\screvise{offset/value}  (\screvise{see Listing 1}) combinations that survives context compaction~\cite{anthropic_compaction}.


\smallskip\noindent
\textbf{Reflection:}
After five consecutive failed tests (i.e., test compiles, does not trigger a crash and there are no new states observed), \FT injects a ``\emph{reflection required}" message. This message directs the agent to inspect mutation count in its test history (to detect \screvise{filters}  that never matched), and rewrite its strategy before running any further tests.


\smallskip\noindent
\textbf{Mutation Taxonomy:}
To prevent the agent from collapsing onto a single class of mutation, \FT\ requires every test to be tagged with a \emph{mutation category} drawn from an eight-class taxonomy. The agent is instructed to track per-category coverage in its strategy memo and is capped at three consecutive tests in the same category. The taxonomy serves two roles: it forces structural diversification during exploration, and it gives the reflection mechanism a concrete axis along which to detect missing strategies. \revise{Please see Appendix~\ref{sec:appendix_mutation} for details regarding the mutation taxonomy.}


\smallskip\noindent
\nsurevise{
\textbf{Seeding Strategies:} Seeding is an essential component for any fuzzing input and the success of the fuzzing campaign
is heavily influenced by the quality of the initial seeds~\cite{seed_selection_for_sucessful_fuzzing,evaluating_fuzz_testing}.
\FT derives its own seeds through its tools and reasoning. The \emph{seeding strategy} for \FT is an alterable component expressed
purely through natural language. In this work, the default seeding strategy is using external knowledge (web search, relevant prior work) to find a suitable initial seed corpus. 
Alternatively, we demonstrate that in the absence of these external knowledge sources, we instruct the \FT
agent to do a broad sweep within a candidate packet. We show that this alternative 
strategy lets \FT function well even with limited access to external knowledge sources
(See Section~\ref{sec:v3-filtered-results}). It is interesting to note that both seeding strategies are just prompts fed to the same agent, so switching between the two strategies requires no code change.
}

\smallskip\noindent
\textbf{Post First Crash Directive:}
On the first confirmed unique crash within a fuzzing session, \FT appends a \emph{FIRST-CRASH-CONFIRMED} directive. This 
directive redirects the agent to focus efforts on discovering new crash variants. Subsequent system prompts also select a post-first-crash variant, 
reinforcing the messaging to focus efforts on discovering new crash variants. 

\subsection{Tool Interface}
\label{sec:tool-interface}
The agent's tool schemas are organised in nine modules. Functionally the tools can be grouped as follows

\begin{figure}[htbp]
    \centering
    \includegraphics[width=\linewidth]{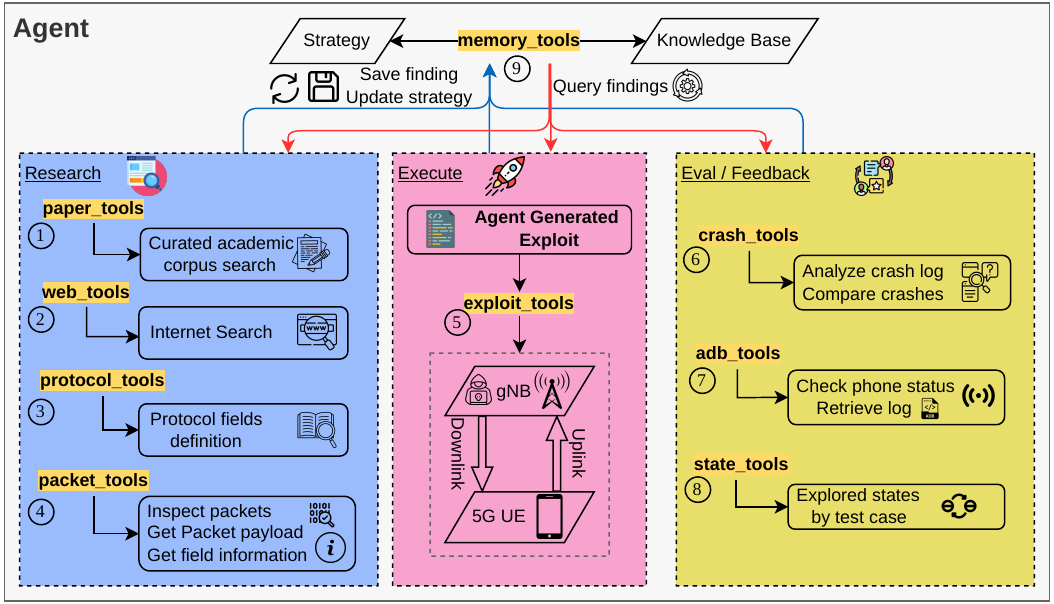}
    \caption{Tools in \FT}
    \label{fig:tool_usage}
\end{figure}

\smallskip\noindent
\textbf{Packet inspection}: The following tools are used to get relevant information from the packet captures: 
\begin{itemize}
\item \texttt{list\_packet\_types}: \grevise{Enumerates unique packet types in a capture file, providing a compact summary of the corresponding filter, state, descriptive information, and occurrence count.}
\item \texttt{inspect\_packets}: \grevise{Retrieves fully dissected packet fields for packets matching a specified filter or index range.}
\item \texttt{get\_packet\_hex}: \grevise{Retrieves the raw hexadecimal bytes of a packet specified by index.}
\item \texttt{get\_field\_info}: \grevise{Retrieves the byte offset, bitmask, and size of a specified protocol field.}
\end{itemize}

Similar tools are used in conventional protocol fuzzing works to 
get packet-level information during fuzzing~\cite{5ghoul}. See marker~\circled{4} in Figure~\ref{fig:tool_usage}. 

\smallskip\noindent
\textbf{Exploit execution}: 
\screvise{While the exploits are systematically generated as test cases  
using the \FT agent, we leverage the infrastructure provided by prior 
works~\cite{5ghoul,U-fuzz} to execute these exploits on smartphones.}
Concretely, the \FT agent moves from a hypothesis to a concrete test against the UE. The tool \texttt{submit\_custom\_exploit} performs a static validation of an exploit. 
Specifically, the tool validates whether the required function signatures 
\grevise{such as \texttt{tx\_post\_dissection} used to mutate packets payload} exist for the underlying \screvise{\FG setup to launch exploits}. 
\grevise{In cases where function signatures are missing or incorrect, this validation fails, and the exploit is therefore not executed.}
The \texttt{execute\_test} tool helps the agent run one exploit file through the underlying setup aforementioned, and similarly with the \texttt{execute\_test\_batch} tool, 
the \FT agent can run up to five tests, which are executed sequentially in a single tool call. The tool call returns a rich structured result: 
test id, compile success/error, execution time, crash details, newly observed states and transitions, capture and log paths, mutation delivery count, a sample of the mutation log, and an inline plain-text analysis that flags common failure modes. The \texttt{execute\_test\_batch} 
tool is only available for the \emph{V3} agent version and above. This module corresponds to marker~\circled{5} in Figure~\ref{fig:tool_usage}.

\smallskip\noindent
\textbf{Crash and State Information}: These tools aid the agent in device-side inspection after a test exploit is run. The \texttt{analyze\_crash\_log} tool parses the device monitor log in a test folder, extracts crash IDs (with timestamps) leveraging the standard 
crash-oracle pattern used in prior 5G fuzzing works~\cite{5ghoul}.
%
Concretely, the tool reads the log monitor and returns the crash context ($\approx$1 KB context slice around the \marker{crash-ID} field). 
The tool \texttt{read\_device\_log} reads lines from a device log file, which is capped at 200 lines per call. 
\grevise{This tool is needed to retrieve recent diagnostic messages from the device log, 
which can support the analysis of crashes and abnormal behavior.} 
Finally, the \texttt{view\_state\_machine} tool builds the full state-transition graph from a pcapng capture leveraging state-transition utilities used in prior works~\cite{U-fuzz,5ghoul}. This returns all observed states and 
transitions so that the agent can reason about protocol-level coverage. See marker \circled{6} and \circled{8} in Figure~\ref{fig:tool_usage}.

\smallskip\noindent
\textbf{Device inspection}: These tools leverage Android Debug Bridge (ADB) to provide agent-side visibility into the UE. The \texttt{adb\_logcat} tool retrieves the last $N$ minutes of Android logs (default 5, capped at 2000 lines) with an optional case-insensitive regex filter. This gives the agent the ability to retrieve targeted insights that the deterministic oracle may have missed. The \texttt{adb\_dmesg} tool reads the kernel ring buffer and is intended for kernel panics, and lower level hardware errors that do not appear in the logcat. The \texttt{adb\_dumpsys} tool retrieves information from (\texttt{telephony.registry}, \texttt{phone}, \texttt{carrier\_config}, \texttt{netd}, \texttt{connectivity}), giving the agent a structured view of the radio and telephony state.  The \texttt{adb\_device\_state} tool returns a consolidated snapshot of the device: baseband version, SIM state, network type, operator, radio technology boot status, uptime, and a subset of \texttt{telephony.registry}. The agent can use it to confirm the modem is alive and attached after a test. The \texttt{adb\_modem\_health} tool is a lightweight liveness probe. See marker \circled{7} in Figure~\ref{fig:tool_usage}.

\smallskip\noindent
\textbf{Memory Tools}: These tools provide the agent access to an explicit knowledge-base that persists beyond a single context window. This means that the observation and strategy can survive context summarisation and can even be persisted across runs. The \texttt{save\_finding} tool records a free-form description with tags (e.g., \emph{crash}, \emph{hypothesis}) and a list of related \marker{test ID}s. This allows the agent to record observations, crash patterns, hypotheses and failures. To aid in retrieval, the agent may use the \texttt{query\_findings} tool to search the observations by tag or keyword. The system also allows the agent to investigate the past test history using the \texttt{get\_test\_history} tool, which includes descriptions, crash status, and any new states or transitions discovered. The \FT agent also maintains a markdown strategy memo that is re-injected into the system prompt at every turn. This aids in maintaining continuity 
and is updated using the \texttt{update\_strategy} tool. This module corresponds to marker~\circled{9} in Figure~\ref{fig:tool_usage}. 

\smallskip\noindent
\textbf{Web Search}: These tools let the \FT agent engage with 
prior work (see marker~\circled{2} in Figure~\ref{fig:tool_usage}). The goal is to let the agent's exploration be guided by known vulnerabilities. The \texttt{web\_search} tool queries a configurable search provider and returns \emph{(title, URL, snippet)} triplets for the top results. The \texttt{fetch\_url} tool then retrieves and extracts plain text from a chosen URL (CVE advisory, blog post, paper page). There are two domain-specific modifications for the security-research setting for the \texttt{fetch\_url} tool: {\em (i)}  GitHub blob URLs are auto-rewritten to \emph{raw.githubusercontent.com} for reliable code extraction. {\em (ii)} GitHub repository roots are resolved through the Contents API into a readable directory listing. This is so that 
the agent can browse exploit repositories rather than reading HTML.  
%

\smallskip\noindent
\textbf{Local Paper Repository Search}: Complementing the web search, 
the \texttt{query\_papers} tool performs a case-insensitive keyword search over the title and abstract of a locally curated \emph{5G} related paper collection with optional \texttt{category} filter. The \texttt{read\_paper} tool then returns the extracted full text of a chosen paper (subject to chunking). The \texttt{list\_paper\_categories} enumerates all the available categories 
\grevise{such as \emph{Wireless Fuzzing and Testing} and \emph{Overshadowing and Injection Attacks}} with paper counts. This gives the agent a map of the corpus before issuing targeted queries. See marker \circled{1} in Figure~\ref{fig:tool_usage}.


\begin{table*}[!t]
\centering
\caption{Modules and Their Associated Tools}
\label{tab:modules_tools}
\begin{tabularx}{\textwidth}{|l|X|}
\hline
\textbf{Module} & \textbf{Tools} \\
\hline
\texttt{packet\_tools.py} & \texttt{list\_packet\_types}, \texttt{inspect\_packets}, \texttt{get\_packet\_hex}, \texttt{get\_field\_info} \\
\hline
\texttt{protocol\_tools.py} & \texttt{search\_fields} \\
\hline
\texttt{state\_tools.py} & \texttt{view\_state\_machine} \\
\hline
\texttt{crash\_tools.py} & \texttt{analyze\_crash\_log}, \texttt{compare\_crashes}, \texttt{read\_device\_log} \\
\hline
\texttt{adb\_tools.py} & \texttt{adb\_logcat}, \texttt{adb\_dmesg}, \texttt{adb\_dumpsys}, \texttt{adb\_device\_state}, \texttt{adb\_modem\_health} \\
\hline
\texttt{exploit\_tools.py} & \texttt{submit\_custom\_exploit}, \texttt{execute\_test}, \texttt{execute\_test\_batch}, \texttt{read\_exploit\_source}, \texttt{list\_exploits} \\
\hline
\texttt{memory\_tools.py} & \texttt{save\_finding}, \texttt{query\_findings}, \texttt{get\_test\_history}, \texttt{update\_strategy}, \texttt{promote\_finding}, \texttt{query\_global\_findings} \\
\hline
\texttt{paper\_tools.py} & \texttt{query\_papers}, \texttt{read\_paper}, \texttt{list\_paper\_categories} \\
\hline
\texttt{web\_tools.py} & \texttt{web\_search}, \texttt{fetch\_url} \\
\hline
\end{tabularx}
\end{table*}

\subsection{Agent Versioning}
\label{anc:agent_version}
\begin{table}
\centering
\caption{\FT version configurations. Each version progressively unlocks capabilities, isolating the contribution of GitHub-aware retrieval (V2) and batched execution (V3). \ngrevise{V3-filtered is a restricted variant of V3.
}}
\label{tab:agent-versions}
\resizebox{\linewidth}{!}{%
\begin{tblr}{
  row{1} = {font=\bfseries},
  column{2-5} = {c},
  hline{1,8} = {-}{1px},
  hline{2-7} = {-}{0.5px},
}
Capability                                 & V1         & V2         & V3         & V3-filtered     \\
Basic tools (e.g., packet inspection tools)& \checkmark & \checkmark & \checkmark & \checkmark      \\
Local paper corpus                         & \checkmark & \checkmark & \checkmark &                 \\
Web search                                 & \checkmark & \checkmark & \checkmark & filtered        \\
GitHub-aware fetch                         &            & \checkmark & \checkmark &                 \\
Batched test execution                     &            &            & \checkmark & \checkmark      \\
\revise{Context summarization threshold}   & 50,000     & 50,000     & 200,000    & 200,000
\end{tblr}
}
\end{table}

To isolate the contributions of each of the subsystems of \FT, we expose \grevise{three} versions of \FT. 
These are runtime configurations of \FT that can be selected by a single flag. A version configuration is a set of boolean flags that gates:
\begin{enumerate}
    \item content to be injected into the system prompt,
    \item tools registered with the LLM powering \FT,
\end{enumerate}

\smallskip \noindent
Table~\ref{tab:agent-versions} summarizes the capability of \FT versions. 

\begin{itemize}
    \item \textbf{V1}:  The agent has access to the protocol-introspection, exploit-execution (single test only), crash and state inspection, device inspection, and the within-session memory tools. For external knowledge, it uses the local paper corpus and the generic web search. When the agent fetches URLs, they are returned as raw HTML. 

    \item \textbf{V2}: V2 extends V1 with the GitHub intelligence layer described in the tool interface (see Section~\ref{sec:tool-interface}). While fetching the URLs, the tool rewrites blob URLs to \emph{raw.githubusercontent.com} and resolves repository roots through the GitHub Contents API into a readable directory listing. This helps \FT to navigate exploit repositories correctly and read source files reliably. 

    \item \textbf{V3}: V3 extends V2 by registering the batch testing tool, allowing the \FT agent to dispatch up to five exploits in a single tool call that are then executed sequentially. \revise{Because the batched execution tool results are larger, we raise the context summarization threshold from {50},{000} to {200},{000} tokens, preventing premature summarization that may discard batch results.}

    \item \textbf{V3-filtered}: \nsurevise{V3-filtered is adapted from V3 by limiting the agent's access to external knowledge. To this end, 
    we disable 
    paper tools and block the access to 
        \textit{github.com}, \textit{gitlab.com}, \textit{gitlab.org}, \textit{githubusercontent.com},
        and \textit{asset-group.github.io}\footnote{we specifically 
        exclude the ASSET Research Group website, as \FG disclosure is hosted on this domain} 
        in web tools. V3-filtered also 
        implements an \emph{alternative broad sweep-based} seeding strategy to continue finding appropriate seeds in the absence of 
        certain external knowledge sources. }

\end{itemize}

\subsection{Execution and Oracle}


\FT leverages prior works~\cite{5ghoul,U-fuzz}, which is an over-the-air (OTA) 
fuzzing framework 
for launching 5G exploits to target devices. 
The exploit development is facilitated through a structured C++ interface, 
which is lightweight and amenable to automatic generation (e.g., by LLMs). 
Such exploits operate primarily at the packet and byte levels, 
and leverage simple APIs to support flexible packet mutation. 
Within an exploit, \FT specifies target packets using Wireshark-style filters and 
directly manipulate packet payloads as needed. 
Once the exploit is ready, it is compiled into a module, which 
applies the specified packet manipulation logic at runtime. 
While an exploit may or may not induce a crash, \FT primarily relies on 
tools analyzing ADB logs from the target device to determine whether a crash has occurred 
given that the target is a smartphone. In the case of non-phone targets, 
\FT instead determines crash by using tools examining device-specific logs (e.g., serial console outputs), if available. 
When such logs are absent or insufficient, potential crashes are inferred by monitoring 
the communication behavior of the target, such as detecting the non-responsiveness. 
In our evaluation, \ngrevise{target devices include phones and BT devices, thus, 
crash detection is performed through the analysis of ADB logs and serial port output.}
In particular, modem crashes are typically accompanied by distinctive error indicators in the logs, 
such as \texttt{Modem Resets!} and \texttt{modem\_failure}, 
\ngrevise{whereas BT devices output error indicators like \texttt{Guru Meditation Error} from serial port}. 
These indicators are not observed during normal, benign communication, making them reliable signals for crash detection. 
For a detailed treatment of the underlying framework, we refer readers to the following resources~\cite{5ghoul,U-fuzz, 5ghoul-github, braktooth}.

The oracle, packet capture handling, and post-processing are deterministic and 
the \FT agent interacts with 
this deterministic system through the exploit execution tools 
(see Section~\ref{sec:tool-interface}). 


\smallskip\noindent
\textbf{Unique Crashes:}
\label{anc:unique_crashes}
We define a crash as \emph{unique} within a \nsurevise{\FT fuzzing} session by its 
crash identifier string (crash ID). This string is extracted from the \ngrevise{device log} by the same crash-oracle pattern similar to prior work~\cite{airbugcatcher}. After every executed test, \FT parses out the crash ID, and checks it against the set of crash IDs already recorded in the session's external knowledge base. \nsurevise{It is important to note that this knowledge base is not 
accessible to the \FT agent, it is just used to make a deterministic presence check.}
If the crash ID is absent, then the crash is recorded as a \emph{new unique crash}, else the test is retained in the test history.

\subsection{\ngrevise{Crash Root Cause Analysis and Mitigation}}
\label{sec:root_cause_and_mitigation}

\ngrevise{
The crashes discovered by \FT were observed on COTS devices running \textit{proprietary} firmware. 
Hence, in the absence of source code, we analyzed all diagnostic information available from the tested devices. 
For example, the ADB logs reported the following assertion failure: 
\textit{[ASSERT] file:mcu/l1/nl1/internal/md97/src/tx/nr\_tx\_mngr\_mcu.c line:817 p1:0x00000000}. 
This information localizes the failure to a specific firmware component 
and source-code location, 
but does not reveal the failed condition or the underlying implementation flaw.
Based on the triggering inputs and the available diagnostic evidence, 
the crashes may result from insufficient validation or improper 
handling of invalid or unexpected values in the firmware. 
However, due to the closed-source firmware, 
it is impossible to highlight if the failure is caused by 
an assertion violation, an invalid state transition, unsafe 
memory access, or another error condition. 
Consequently, we cannot prescribe or validate exact 
source-level mitigation measures for the affected firmware.}

\scnrevise{As security researchers, we have responsibly disclosed 
the crash-inducing inputs to the affected vendors and collaborated 
with them to investigate potential mitigation measures and firmware 
patches. In general, several discovered failures could be mitigated 
in the firmware by strengthening the validation of externally 
controlled values before processing them and by introducing graceful 
error-handling and recovery mechanisms for unexpected inputs.}

%

\section{Case Study}
\label{sec:case-studies}
\subsection{Case Study: An Actual \FT Run}


In this case study, we demonstrate how \FT operates, 
focusing on the structure and formulation of prompts across multiple interaction turns, 
as well as the tools invoked during execution. 
The study is conducted using \FT V3 with a OnePlus Nord CE 2 5G target. 
Prior to the experiment, a set of tools is prepared, as summarized in Table~\ref{tab:modules_tools}, 
along with several prompts. We note that such is a one-time effort. 
The experiment begins with the initial prompt (Prompt 1), which is sent to the configured LLM in Turn 1. 
A simplified version of this initial prompt is shown in Listing~\ref{lst:prompt_1}.
In this prompt, we assign the role of 5G expert to LLM, describe the test environment, define the task objective 
and provide details of the available tools that LLM could understand and invoke. 
In response, the LLM produces Response 1 as shown in Listing~\ref{lst:response_1}, 
which includes an analysis of the fuzzing task and identifies a set of tool calls. 
The response demonstrates that LLM starts by conducting researches about 
known vulnerabilities before running any exploits. Specifically, the LLM wants to invoke two 
tool calls in Response 1: \textit{web\_search} and \textit{query\_papers}.
In the next step, \FT executes the tools as specified in Response 1, 
and their outputs are collected for use in the next turn. 
The LLM request, LLM response and tool execution form a complete turn.
Subsequently in Turn 2, the request and response in Turn 1, together with the tool call results, 
are concatenated to form Prompt 2, as shown in Listing~\ref{lst:prompt_2}. 
Specifically, the tool calls are tracked by \texttt{tool\_call\_id} during 
the experiment.  This  enables LLMs to link a tool call and its result. 
The LLM generates a response and identifies the tools to be executed. 
The results are then collected, and the process repeats, forming an iterative interaction loop. 

\lstdefinestyle{prompt}{
  basicstyle=\footnotesize\ttfamily,
  float,
  breaklines=true,
  breakindent=0pt,
  breakautoindent=false,
  backgroundcolor=\color{gray!15},
  frame=single,
  framerule=0.5pt,
  rulecolor=\color{gray!150},
  framesep=5pt,
  xleftmargin=5pt,   
  xrightmargin=5pt,   
  captionpos=b
}

\begin{lstlisting}[
  style=prompt,
  caption={Prompt 1},
  label={lst:prompt_1}
]
You are a 5G NR security researcher conducting over-the-air (OTA) fuzz testing against a OnePlus UE (User Equipment) running a MediaTek modem using the WDissector framework. 

Your goal is to discover unique modem crashes by crafting and executing targeted protocol mutations.

## Workflow
1. Inspect
2. Research (REQUIRED):
   a. Online exploit search (REQUIRED)
   b. Academic research
   c. For each paper or CVE, extract and record the details via `save_finding`
3. Get a crash first (TOP PRIORITY)
4. Hypothesize: Based on protocol knowledge and research, identify fields likely to trigger modem bugs
5. Test: Write a C exploit and run it
6. Analyze: Read the `analysis` field in the result
7. Record (MANDATORY): You MUST record what you tried, what happened and what you learned.
8. Update strategy: You need to maintain your fuzzing strategy
9. Iterate: Adjust strategy based on results. Explore related mutations near crashes

## Available Tools
- web_search: Search for exploit write-ups, vulnerability reports, security advisories, and technical articles
  Input: query: string, max_results: int
- query_papers: Search stored papers in the local 5G paper collection
  Input: keyword: string, max_results: int
- ...
\end{lstlisting}

\begin{lstlisting}[
  style=prompt,
  caption={Response 1},
  label={lst:response_1}
]
I'll start by surveying the packet types, researching known vulnerabilities, and then designing targeted tests to discover crashes. Let me begin systematically. 

I want to make the following tool calls:
- tool_call_id: 1, web_search({"query": "MediaTek modem 5G NR RRC crash vulnerability CVE", "max_results": 10})
- tool_call_id: 2, query_papers({"keyword": "fuzzing baseband modem", "max_results": 10})
\end{lstlisting}

\begin{lstlisting}[
  style=prompt,
  caption={Prompt 2},
  label={lst:prompt_2}
]
{Prompt 1}
{Response 1}

## Tool call results
- tool_call_id: 1 
  1. title: Vulnerability in Qualcomm and MediaTek 5G modems affects..., url: ...
  2. title: New attack impacts 5G phones with Qualcomm, MediaTek..., url: ...
  ...
- tool_call_id: 2
  1. paper: 5G Fuzzing, abstract: Effective fuzzing framework to find vulnerabilities...
  2. paper: Baseband Bugs, abstract: We propose a systematic way to locate the malicious...
  ...
\end{lstlisting}

According to our observations, the \FT agent primarily focused on survey and information gathering 
during the initial turns. In this case study, the agent devoted Turns 1--14 to research activities, 
systematically recording intermediate discoveries using the \textit{save\_finding} tool. 
These findings were stored in a knowledge base that the agent could 
subsequently query and reference through retrieval tools. 

Beginning in Turn 15, the agent transitioned from survey to exploit generation and execution 
by leveraging the \texttt{execute\_test\_batch} tool (see  Table~\ref{tab:modules_tools}). Concretely, the agent 
generated and executed multiple exploit scripts, several of which successfully triggered crashes. 
After each exploit execution, the \FT agent also invoked 
\texttt{save\_finding} and \texttt{update\_strategy} tools to document execution outcomes and iteratively 
refine its fuzzing strategy. This feedback loop enabled the agent to adapt its 
subsequent actions based on prior experimental results. 

After discovering the first crash, the agent dynamically modified its operational objective 
to prioritize the discovery of \textbf{novel} crashes. 
Specifically, \FT updated \textit{Workflow Point 3} in the prompt shown in Listing~\ref{lst:prompt_1} 
to: \texttt{NOVEL CRASH HUNTING: Your toolchain is proven. Your SOLE objective now is finding novel crashes}. 
The updated prompt was then applied in all subsequent turns. 
Following this objective shift, \FT continued researching, generating and executing exploits 
until the configured turn budget is reached. 

\subsection{Case Study: Building \revise{the \FT} Agent}
\label{sec:case_study_build_agent}

\noindent
\textbf{GitHub Content Retrieval Optimization:} 
We followed an iterative process when developing \FT. 
After making changes to \FT, such as modifying tools and refining prompts, we evaluated 
whether \FT was still functioning effectively. 
Specifically, we examined the LLM prompts and responses to identify any anomalies and 
assessed whether there were any changes in fuzzing performance. 
During the development of the \texttt{fetch\_url} tool, we observed that it returned redundant 
HTML structures when processing GitHub repository URLs. 
We examined the tool outputs by inspecting execution logs, which revealed that they often contained 
substantial amounts of irrelevant content including HTML structural tags, CSS styles or JavaScript code. 
This redundancy increased token consumption and reduced the efficiency of LLM analysis. 
To address this, we enhanced the tool by integrating GitHub APIs to retrieve the repository 
tree structure in clean JSON format when the input corresponds to 
a GitHub repository URL. 
This optimization can be extended to other web URLs. However, we specifically optimized 
for GitHub URLs, as we observed a high frequency of calls to them. 

\noindent
\textbf{Batching Exploit Execution:}
The initial \FT system only provided the \textit{execute\_test} tool, 
which allowed \FT to execute a single exploit per tool call. 
This design introduced overhead when evaluating multiple candidate exploits, 
as the LLM was required to generate separate tool calls for each execution attempt. 
Consequently, the cost and complexity of repeated tool invocation discouraged broader exploit exploration. 
To address this limitation, we developed the \textit{execute\_test\_batch} tool, 
which enables \FT to enqueue and execute multiple exploits sequentially within a single tool call. 
This batching mechanism reduces tool-calling overhead and encourages more comprehensive exploit evaluation. 
\grevise{Section~\ref{sec:rq_effective} demonstrates the performance improvement achieved by employing this tool in the V3 configuration.}


\noindent
\textbf{Research Quality:}
\label{anc:rq}
While developing \FT, we observed that the agent autonomously performs 
a preliminary research phase before fuzzing. This raises the question of 
how to assess the quality of this research. In practice, low-quality research 
may return irrelevant information (e.g., unrelated papers or websites), 
which can reduce fuzzing effectiveness and require improvements to the research process or tools.

\begin{table}[b]
\centering
\caption{Research quality \grevise{scores}} 
\label{tab:rq_level}
\resizebox{\linewidth}{!}{%
\begin{tblr}{
  row{1} = {font=\bfseries},
  column{1} = {c},
  cell{2}{1} = {font=\bfseries},
  cell{3}{1} = {font=\bfseries},
  cell{4}{1} = {font=\bfseries},
  cell{5}{1} = {font=\bfseries},
  cell{6}{1} = {font=\bfseries},
  cell{7}{1} = {font=\bfseries},
  hline{1-2,8} = {-}{},
  hline{1,Z} = {1pt},
}
Score & The agent...                                                                                                   \\
0     & Provides no relevant vulnerability-related information                                                         \\
1     & Mentions a related project or known vulnerability (e.g. ``CVE-2023-32842'')                                        \\
2     & Correctly identifies the relevant message type (e.g., ``RRC Setup'')                                           \\
3     & Identifies the relevant substructure within the message \grevise{(e.g., ``searchSpace'')}                      \\
4     & Provides precise byte-level details (e.g., offset and value)                                                   \\
5     & {Triggers a crash for the specific input instance using the retrieved \\information, with score 4-level knowledge demonstrated} 
\end{tblr}
}
\end{table}
\begin{table*}
\centering
\caption{Test Device Specifications}
\label{tab:test_device}
\resizebox{\linewidth}{!}{%
\begin{tblr}{
  row{1} = {font=\bfseries},
  hline{2,7} = {-}{},
  hline{1,Z} = {1pt},
}
Protocol & Device Model         & Modem                   & Firmware                     & Android Security Patch & Device ID       \\
5G       & OnePlus Nord CE 2 5G & MediaTek Dimensity 900  & IV2201\_11\_F.48             & 2023-05-05             & \texttt{mediatek}        \\
5G       & REDMI 15C 5G         & MediaTek Dimensity 6300 & AP3A.240905.015.A2           & 2025-08-01             & \texttt{mediatek\_d6300} \\
5G       & OnePlus Nord 3       & MediaTek Dimensity 9000 & CPH2493\_16.0.3.500(EX01)    & 2026-02-01             & \texttt{mediatek\_d9000} \\
5G       & OnePlus Ace 5 Ultra  & MediaTek Dimensity 9400 & PLC110\_15.0.2.610(CN01)     & 2025-09-01             & \texttt{mediatek\_d9400} \\
5G       & Hisense F50+         & UNISOC UD710            & PPR1.180610.011 release-keys & 2020-12-05             & \texttt{unisoc\_ud710}   \\
BT & ESP-WROOM-32         & N.A.                    & ESP-IDF commit 3de8b79       & N.A.                   & \texttt{esp32}           
\end{tblr}
}
\end{table*}

To evaluate the usefulness of 
the agent's research, we introduce the metric 
\textit{research quality}, which measures how well the retrieved information 
supports effective fuzzing. The textual research output
is automatically evaluated using predefined rules. 
We define a research quality scoring scheme ranging from 0 (lowest) to 5 (highest), 
as shown in Table~\ref{tab:rq_level}. These scores 
capture increasing relevance and precision of vulnerability-related knowledge. 
Evaluation is performed in a top-down manner: the system first checks whether 
the criteria for score 5 are met and otherwise proceeds to lower scores. 
Lower scores reflect coarse understanding (e.g., referencing projects or CVEs), 
while higher scores capture finer-grained technical details e.g., protocol 
structures and byte-level information. At score 5, the research enables a 
crash during fuzzing, indicating that the retrieved knowledge is both accurate 
and actionable.

The scores are not strictly cumulative; instead, the metric primarily serves as a diagnostic tool 
to assess whether the agent conducts meaningful and actionable research. 
Currently, \textit{research quality} is evaluated only on the \texttt{mediatek} (MediaTek Dimensity 900) target, 
where public vulnerability and crash-related resources enable objective assessment 
(e.g., mapping to known CVEs). 
For other targets, the lack of ground truth makes these criteria difficult to apply. 

In particular, achieving score 5 requires that a crash is triggered during the fuzzing session, 
and that the agent has already demonstrated score 4-level knowledge, ensuring that the success 
is grounded in sufficiently detailed and actionable understanding rather than incidental discovery.

\section{Evaluation Setup}
\label{sec:eval-setup}

\smallskip\noindent
\textbf{Hardware Setup:} 
We evaluate \FT on \ngrevise{two protocols: Bluetooth Classic (BT) and 5G. The tested 5G devices include five phones} from three manufacturers, equipped with modems from two major 5G chipset vendors: MediaTek and UNISOC. 
\grevise{According to Counterpoint Research~\cite{counterpoint2026smartphoneapsoc}, 
MediaTek and UNISOC accounted for 32\% and 14\%, respectively, of global smartphone AP/SoC shipments in Q1 2026.}
\ngrevise{For Bluetooth experiments, we used ESP32-Ethernet-Kit as the host board to establish 
connection with target ESP-WROOM-32. Instructions for setting up the host and target boards are provided in prior work~\cite{braktooth}. 
}
\ngrevise{The 5G devices} were selected to represent the different modem implementations 
as well as the differences between smartphone brands. 
Table~\ref{tab:test_device} summarizes the devices evaluated (hereafter referred to as \textit{targets}). 
\ngrevise{The \texttt{Firmware} column reports the firmware version associated with each target. 
For 5G targets, this information is obtained using the command \texttt{adb shell getprop gsm.version.baseband}, 
whereas for BT targets, it refers to the ESP-IDF commit of the flashed firmware.}
For clarity, we use the \textit{Device ID} defined in Table~\ref{tab:test_device} to 
consistently refer to each target throughout the text and the evaluation. 

Each 5G testbed consists of a host PC and a software-defined radio (SDR). 
The two testbeds are situated in different rooms within the office to avoid potential interference. 
The PC runs both the \FG stack and \FT, while the SDR facilitates 5G NR communication with the targets. 
One testbed is powered by an AMD Ryzen AI Max+ 395 processor with 128~GB of RAM, 
whereas the other uses an AMD Ryzen 7 5800H processor with 12~GB of RAM. 
Both systems run Ubuntu 22.04.5 LTS and are equipped with USRP B210 SDRs. 


\smallskip\noindent
\textbf{Software Setup:} 
\nsurevise{\FT is implemented in Python (Approximately 16,070 LoC covering agent logic, tool implementations, and tests).
We used Claude Code~\cite{claude_code} to assist with code generation and an existing 5G fuzzing framework~\cite{5ghoul-github} to
launch exploit test cases.
}

\smallskip\noindent
\textbf{Choice of LLMs:}
\nsurevise{We evaluated \FT on six LLMs, balancing evaluation coverage against cost of experiments. We were guided by 
existing LLM leaderboards~\cite{llm_stats}. The criteria for selection were: \emph{capability} (inclusion of flagship SOTA models), 
diversity (models from multiple LLM providers), and open-source practicality (to compare open source models to proprietary ones).
The resulting set of models (See Table~\ref{tab:models}) contains four widely adopted commercial models and two state-of-the-art open-source models. All of the models were released in 2026. In our evaluations, we adopt the OpenAI API format for API invocation using OpenRouter~\cite{openrouter} and Microsoft Foundry~\cite{azureaifoundry}. 
}
%
%

\begin{table}[htbp]
\centering
\caption{Summary of evaluated LLMs, with cost in USD per million tokens.}
\label{tab:models}
\resizebox{\linewidth}{!}{%
\begin{tblr}{
  row{1} = {font=\bfseries},
  cell{2}{4} = {r=4}{},
  cell{6}{4} = {r=2}{},
  hline{1-2,6,8} = {-}{},
  hline{1,8} = {1.2pt},
}
Model             & Provider   & Release  & License     & Cost (Input / Output) \\
Claude Sonnet 4.6 & Anthropic  & Feb 2026 & Proprietary & \$3 / \$15            \\
GPT-5.5           & OpenAI     & Apr 2026 &             & \$5 / \$30            \\
Gemini 3.1 Pro    & Google     & Feb 2026 &             & \$2 / \$12 *          \\
Qwen3.7 Plus      & Alibaba    & Jun 2026 &             & \$0.4 / \$1.6         \\
DeepSeek V4 Pro   & DeepSeek   & Apr 2026 & Open-source & \$0.435 / \$0.87      \\
Kimi K2.6         & MoonshotAI & Apr 2026 &             & \$0.74 / \$3.5        
\end{tblr}
}
\begin{flushleft}
\footnotesize * Gemini 3.1 Pro charges \$4 (Input) and \$18 (Output) per million tokens for prompts exceeding 200K tokens.
\end{flushleft}
\end{table}

\smallskip\noindent
\textbf{Repeated Runs and Analysis:}
\nsurevise{To account for variability introduced by stochastic LLM outputs and fuzzing processes, 
we repeat each experiment five times under 
identical configurations and analyze the aggregated results across all runs.}

\section{Evaluation Results}
\label{sec:eval-results}


\subsection{\revise{How effectively does \FT discover crashes?}}
\label{sec:rq_effective}

\begin{figure}[htbp]
    \centering
    \includegraphics[width=\linewidth]{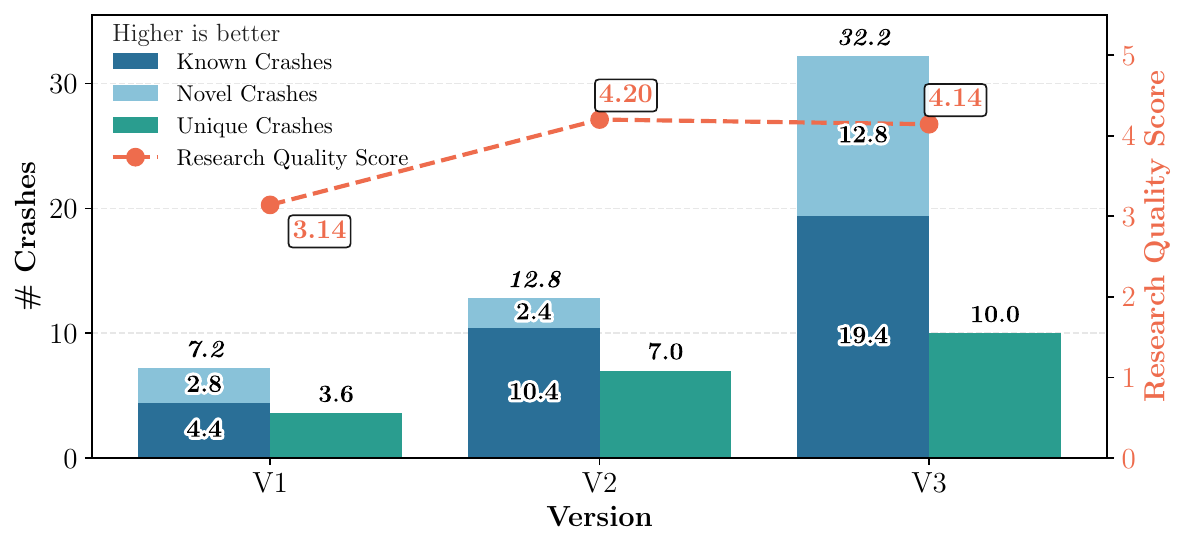}
    \caption{\grevise{\FT performance in discovering crashes under different configurations. 
All reported values are averaged over five runs: the bars show the number of crashes discovered, 
the italic numbers indicate the total number of crashes, and the dashed line represents the research quality score.}
    }
    \label{fig:rq1_crash}
\end{figure}

\mnrevise{To evaluate the effectiveness of \FT, we ran different configurations of system prompts and registered tools}, as described in Section~\ref{anc:agent_version}, using \claude on the \texttt{mediatek} (MediaTek Dimensity 900) target (\mnrevise{see Table~\ref{tab:test_device}}).

Figure~\ref{fig:rq1_crash} presents the discovered crash statistics and research quality scores. 
We categorize crashes that can be found in public resources, such as websites or CVE disclosures, \ngrevise{with available exploit scripts}, 
as \textit{\ngrevise{known}} crashes, while those that do not appear in any publicly accessible sources 
are categorized as \textit{\ngrevise{novel}} crashes. 

Crashes are usually associated with some 
form of identifier~\cite{airbugcatcher}, 
allowing us to count the number of unique crashes. 
\nsurevise{For example, the disclosure page for related work~\cite{5ghoul, 5ghoul-github}} 
\ngrevise{lists many crash identifiers\footnote{such as 
``\textit{[ASSERT] file:mcu/l1/mml1/mml1\_endc/src/mml1\_endc\_db\_hdlr.c line:524 p1:0x91920c70}''} for \texttt{mediatek}. 
Moreover, we have provided a list of known \nsurevise{crash strings} for \texttt{mediatek} \nsurevise{(See Table~\ref{tab:public-crash-ids} in Appendix~\ref{app:public-crash-id}).}}


The results show that introducing additional tools in V2 and V3, 
such as the GitHub fetch tool and the batch testing tool \grevise{as mentioned in Section~\ref{sec:case_study_build_agent}},
effectively improved both crash discovery performance and research quality. 
\grevise{Specifically, the average total number of crashes across five runs increased 
from \textbf{7.2} in V1 to \textbf{12.8} and \textbf{32.2} in V2 and V3, respectively.}
Although the number of \ngrevise{novel} crashes in V2 slightly decreased by 0.4 compared with V1, 
the number of \ngrevise{known} crashes increased substantially to 10.4, representing a \textbf{2.36× improvement}. 
This improvement is likely attributable to the GitHub fetch tool, which enables \revise{enhanced} retrieval of public exploit scripts. 
The batch testing tool introduced in V3, 
\revise{along with its enlarged context window 
(see Section~\ref{anc:agent_version})}, 
led to a significant advancement in discovered crashes, 
with increases of 10 \ngrevise{novel} crashes and 15 \ngrevise{known} crashes \ngrevise{compared to V1}. 
\grevise{Moreover, the average number of unique crashes across five runs 
increased from \textbf{3.6} in V1 to \textbf{7.0} and \textbf{10.0} in V2 and V3, respectively.}
Finally, the average research quality score (see Section~\ref{anc:rq}), computed across five runs, 
\ngrevise{increased from \textbf{3.14} for V1 to \textbf{4.20} for V2, 
before lowering slightly to \textbf{4.14} for V3.}


\begin{center}
\begin{tcolorbox}[width=\columnwidth, standard jigsaw, opacityback=10, arc=0pt, auto outer arc]
\textbf{Key Finding:} \grevise{\textbf{\FT is effective in discovering crashes.}
Compared to V1, V3 achieves a \textbf{4.47$\times$} improvement in the average number of total crashes, 
a \textbf{2.78$\times$} improvement in the average number of unique crashes, 
and a \textbf{\ngrevise{31.8}\%} increase in the average research quality score.}
\\ \\
\grevise{
This demonstrates that GitHub fetch and batch testing tools substantially enhance both crash discovery performance and research quality.}
\end{tcolorbox}
\end{center}

\subsection{\revise{How effectively does \FT discover crashes across different modems?}}
To assess the \grevise{generality} and performance of \FT across different devices, 
we ran \FT V3 against the five targets (see Table~\ref{tab:test_device}), 
with each experiment repeated five times. 

\begin{figure}[b]
    \centering
    \includegraphics[width=\linewidth]{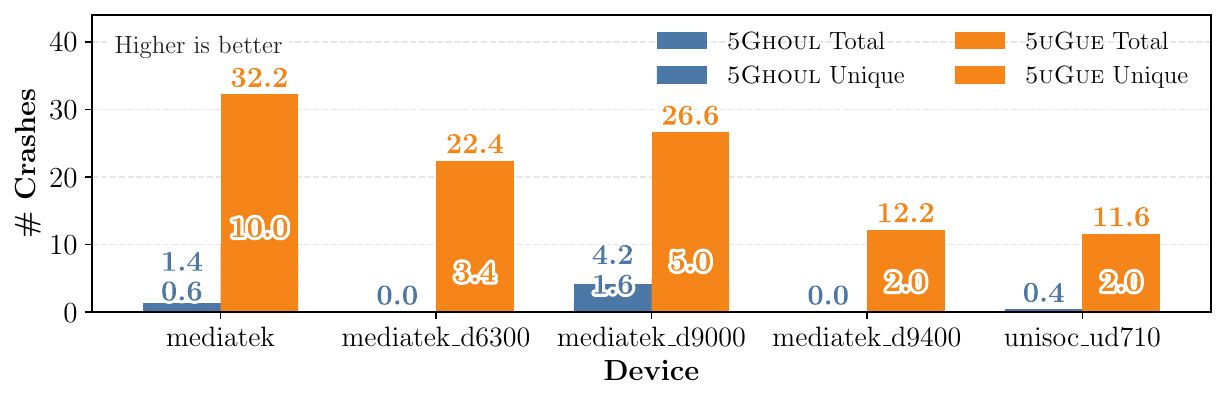}
    \caption{Performance of \FT V3 and \FG in discovering crashes across five targets. 
    For each target, results are averaged over five runs. 
    \textbf{Total} indicates the mean number of crashes per run, 
    while \textbf{Unique} denotes the mean number of unique crashes as defined in Section~\ref{anc:unique_crashes}.}
    \label{fig:compare_crash}
\end{figure}

Figure~\ref{fig:compare_crash} shows the discovered crash statistics. 
The results show that, across the five targets, \FT discovers at least 11 crashes on average, including at least 2 unique crashes. 
Specifically, \FT discovers the highest number of crashes on the \texttt{mediatek} (MediaTek Dimensity 900) target, 
with averages of 32.2 total crashes and 10 unique crashes. 
This is expected, as substantially more online information is available for this target, 
allowing \FT to learn from existing exploit-related knowledge. 
Among the remaining four targets, for which publicly available information is limited, 
\FT is still able to effectively identify more than 10 crashes per target. 
Specifically, \FT discovers the fewest crashes on \texttt{unisoc\_ud710}, 
which may be attributed to the particularly scarce publicly available information about UNISOC chipsets.

Figure~\ref{fig:compare_fuzz_time} presents the actual runtime of \FT. 
The average runtime for each target remains below 105 minutes within 60 turns under the V3 configuration. 
We observe that the runtime varies across the five runs for each target, 
likely due to the agent dynamically selecting different fuzzing paths during execution.

\begin{center}
\begin{tcolorbox}[width=\columnwidth, standard jigsaw, opacityback=10, arc=0pt, auto outer arc]
\textbf{Key Finding:} \grevise{\FT generalizes effectively across different modem targets, 
discovering crashes on all five evaluated targets. 
Averaged over five runs, \FT discovers \textbf{21} total crashes and 
\textbf{4.48} unique crashes per target. 
Across targets with varying levels of publicly available information, 
\FT identifies more than \textbf{11} crashes on every target, 
with total crashes ranging from \textbf{11.6} on \texttt{unisoc\_ud710} 
to \textbf{32.2} on \texttt{mediatek}.}

\end{tcolorbox} 
\end{center}

\subsection{\ngrevise{Does \FT extend to other protocols?}}
\label{sec:bt_result}

\ngrevise{ 
\nsurevise{\FT's agentic loop makes no wireless protocol specific assumptions. 
To demonstrate this, we adapted \FT to Bluetooth Classic (BT) and 
conducted experiments (using the \FT V3 configuration)
on the BT target \texttt{esp32}, making only minimal changes to \FT. 
Specifically, we modified the existing
exploit tools to support launching exploits against BT targets using prior work~\cite{braktooth}.}
We also created a new prompt for the BT target 
by adapting the existing 5G prompts. 
\ngrevise{Appendix~\ref{app:bt_prompt} highlights the key changes made when adapting the 5G prompts for BT fuzzing.}
These changes were basic and procedural, including references to the BT
experimental environment and API documentation for BT exploit generation.}

\ngrevise{
The target BT device, \texttt{esp32} exposes a crash identifier 
through its serial port, which can be used to distinguish unique crashes. 
However, there are no publicly accessible sources documenting 
BT exploits and crash identifiers for \texttt{esp32}. 
Therefore, we categorize all crashes discovered by \FT on 
\texttt{esp32} as \emph{novel crashes}.}

 \ngrevise{
We repeat the experiment five times using \claude. On average, \FT discovered \textbf{6.4} novel crashes,
\textbf{8} total crashes, and \textbf{1.6} unique crashes across the five runs.
The number of novel crashes differs from the total number of crashes because some crashes discovered by \FT
were unclassified, meaning that they did not have a crash identifier
(see Appendix~\ref{app:detailed_results} for details).}
%
%
%
\nsurevise{We observe that \ngrevise{three out of eight} unique BT crashes could not be reproduced, 
an issue also reported in prior work~\cite{airbugcatcher}.
We consider addressing these BT specific limitations beyond the scope of this work.}

\nsurevise{ Overall, our results show that \FT extends to other protocols with 
minimal changes and still triggers crashes without protocol-specific optimization.
}

\begin{center}
\begin{tcolorbox}[width=\columnwidth, standard jigsaw, opacityback=10, arc=0pt, auto outer arc]
\ngrevise{
\textbf{Key Finding:} With minimal procedural changes and no protocol-specific optimization, 
\FT was successfully extended to BT and discovered an average of \textbf{6.4} novel crashes and 
\textbf{1.6} unique crashes on \texttt{esp32} across five runs.
}
\end{tcolorbox} 
\end{center}

\subsection{\ngrevise{How does \FT perform with restricted access to external knowledge?}}
\label{sec:v3-filtered-results}


\begin{figure}[htbp]
    \centering
    \includegraphics[width=\linewidth]{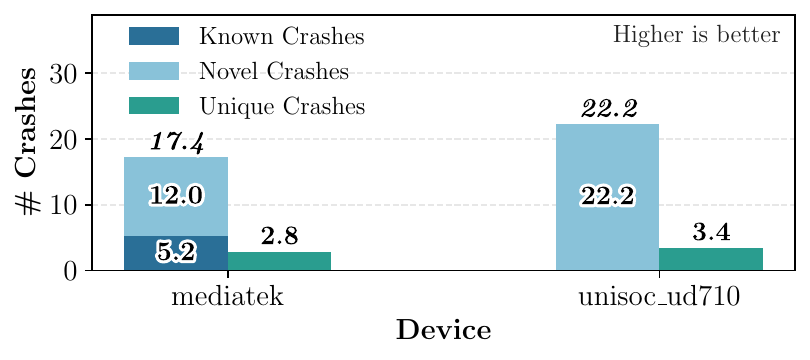}
    \caption{Performance of \FT V3-filtered in discovering crashes in \texttt{mediatek} and \texttt{unisoc\_ud710} targets.
All values are averaged over five runs. The bars represent the number of crashes discovered, 
and the italic numbers indicate the total number of crashes. 
\\$*$ The sum of \ngrevise{known and novel} crashes is not equal to the number of total crashes because 
of an unclassified crash (a crash without crash identifier) for \texttt{mediatek}. 
See Appendix~\ref{app:detailed_results} for details.}
    \label{fig:v3f_crash}
\end{figure}

\nsurevise{
In the standard versions \ngrevise{(V1, V2 and V3)} of \FT, seeding is not 
performed by supplying a fixed corpus of packets or offsets 
(\scnrevise{see Figure~\ref{fig:5g_test_case}}). 
Instead, \FT dynamically derives the initial seed inputs 
(candidate packets, payloads, and mutation targets) by combining externally 
obtained information with the agentic feedback loop.
To test the flexibility of \FT, we conducted 
experiments using \textbf{V3-filtered}, introduced in 
Section~\ref{anc:agent_version}, on two modem targets from different vendors. The 
results are shown in Figure~\ref{fig:v3f_crash}.}
%


\nsurevise{With a seeding strategy identical to V3, V3-filtered could \emph{barely}
discover any crashes i.e., only one of the five runs triggered a crash. 
By analyzing the 
fuzzing strategy in that successful run, we found that 
key move was to \emph{initially} test multiple mutation offsets within 
the same test case, rather than testing individual offsets across 
different test cases.}

\nsurevise{
This motivated an offset-first payload-sweeping strategy and the re-sweeping 
of packets that had previously exposed crashes. 
We instructed \FT to generate 
more test cases in each batch to increase its exploration throughput. }
\ngrevise{Additionally, we 
introduced \textbf{structured crash-map state tracking} which records 
the mutation offsets explored and the corresponding exploit outcomes
allowing \FT to retain the results of previous attempts and avoid redundant exploration.}
%
%

\nsurevise{
Specifically, across five runs, \FT V3-filtered discovers an 
average of \ngrevise{17.4} total crashes (12 novel, \scnrevise{2.8 unique}) 
on \texttt{mediatek},
and an average of 22.2 total crashes (all novel, \scnrevise{3.4 unique}) 
on \texttt{unisoc\_ud710}.
Two of the five runs on \texttt{mediatek} did not discover any crashes. 
Due to the stochasticity in the fuzzing process, such a variation is unsurprising.}

\ngrevise{Overall, these results show that \nsurevise{altering} the seeding strategy 
does not hinder \FT's fuzzing capability. External knowledge facilitates seeding by 
providing useful hints about promising packets, payloads, and mutation targets. 
When such knowledge is limited, \FT can still bootstrap the initial seeding process. 
V3 and V3-filtered differ primarily in their seeding strategies: 
non-filtered versions of \FT derive seeds and mutation targets mainly from external knowledge, 
whereas V3-filtered uses the refined prompt to bootstrap exploration 
without target-specific packets or offsets.\footnote{See Appendix~\ref{app:v3f-prompt} for the refined V3-filtered prompt.}
More broadly, the results demonstrate that \FT's seeding strategy can be readily adapted through natural-language instructions.}

\begin{center}
\begin{tcolorbox}[width=\columnwidth, standard jigsaw, opacityback=10, arc=0pt, auto outer arc]
\nsurevise{
\textbf{Key Finding:} Despite having limited access to external knowledge sources, 
\FT V3-filtered discovers an average of \ngrevise{17.4} and 22.2 total crashes, 
including 2.8 and 3.4 unique crashes, on \texttt{mediatek} and \texttt{unisoc\_ud710}, respectively. 
}
\end{tcolorbox} 
\end{center}

\subsection{How does the choice of LLM affect the performance of \FT?}

\begin{figure}[htbp]
    \centering
    \includegraphics[width=\linewidth]{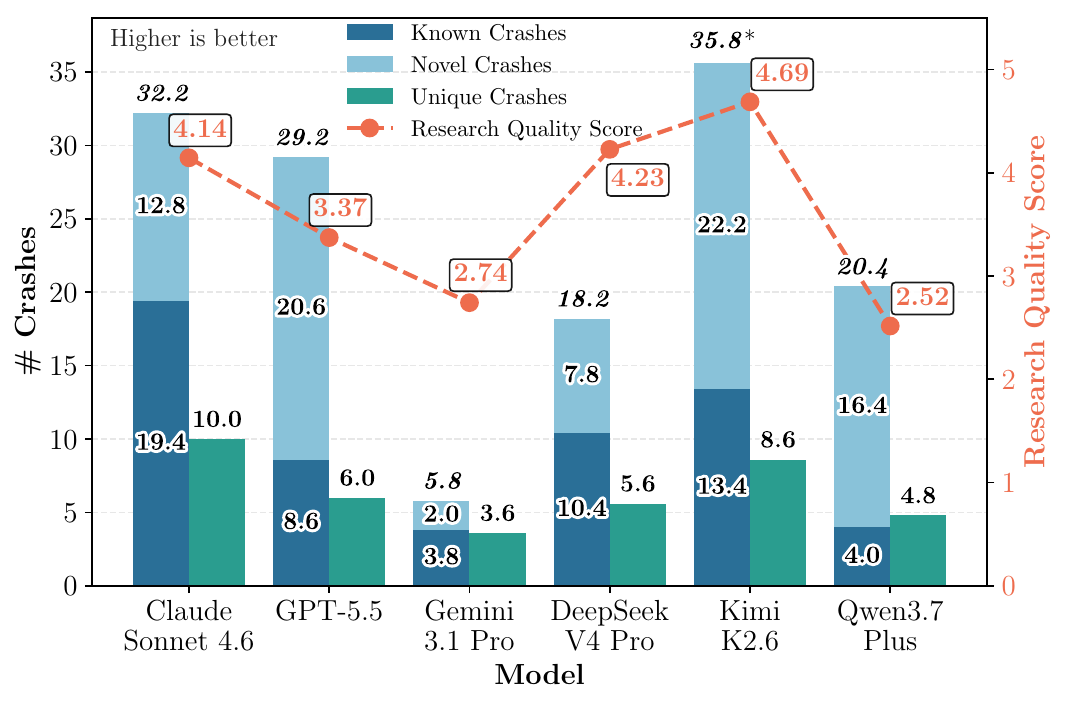}
    \caption{
    \FT's crash discovery performance with different LLMs.
All values are averaged over five runs. The bars represent the number of crashes discovered, 
the italic numbers indicate the total number of crashes, and the dashed line represents the research quality score.
\\$*$ The sum of \ngrevise{known and novel} crashes is not equal to the number of total crashes because 
of an unclassified crash (a crash without crash identifier) for \kimi. 
See Appendix~\ref{app:detailed_results} for details.
    }
    \label{fig:rq2_crash}
\end{figure}

To evaluate the impact of different LLMs on the performance of \FT, 
we conducted tests using six different models with the V3 configuration on the \texttt{mediatek} (MediaTek Dimensity 900) target 
(see Table~\ref{tab:test_device}). 
Figure~\ref{fig:rq2_crash} shows the metrics for discovered crashes and research quality.

\grevise{The results show that \kimi achieves the best overall crash discovery performance among the six models, 
discovering on average \textbf{13.4} \ngrevise{known} crashes, \textbf{22.2} \ngrevise{novel} crashes, 
and \textbf{8.6} unique crashes across five runs. 
\claude ranks second overall, discovering on average \textbf{19.4} \ngrevise{known} crashes, 
\textbf{12.8} \ngrevise{novel} crashes, and the highest number of unique crashes, \textbf{10.0}, across five runs. 
In contrast, \gemini performs the worst, discovering on average only \textbf{3.8} \ngrevise{known} crashes, 
\textbf{2.0} \ngrevise{novel} crashes, and \textbf{3.6} unique crashes across five runs.}

\grevise{
We have also computed the Pearson correlation between research quality score and 
the total number of crashes. 
For each model, the total number of crashes was extracted from Figure~\ref{fig:rq2_crash} 
and paired with the corresponding research quality score. 
The resulting correlation coefficient was \ngrevise{$r = 0.662$}, suggesting a moderate positive correlation~\cite{mukaka2012statistics}.}

Specifically, \gemini and \qwen achieve relatively low research quality scores of \textbf{\ngrevise{2.74}} and \textbf{\ngrevise{2.52}}, respectively, 
and their crash discovery performance ranks among the lowest across the six models. 
In contrast, \kimi achieves the highest research quality score of \textbf{\ngrevise{4.69}},
which is consistent with its strong crash discovery performance. 
Although \ds obtains the second-highest research quality score of \textbf{\ngrevise{4.23}}, 
its crash discovery performance ranks second lowest among the six models. 

\begin{center}
\begin{tcolorbox}[width=\columnwidth, standard jigsaw, opacityback=10, arc=0pt, auto outer arc]
\textbf{Key Finding:} \grevise{The choice of LLM substantially affects the performance of \FT: 
\kimi discovers the most crashes, with an average of \textbf{\ngrevise{35.8}} total crashes, 
which is \textbf{\ngrevise{6.2}$\times$} as that of \gemini, the worst-performing model with only 
\textbf{5.8} total crashes. 
\kimi also outperforms the second-best model, \claude, by \textbf{\ngrevise{11.2}\%} in total crash discovery.}
\end{tcolorbox} 
\end{center}

\subsection{How does \FT compare to the state-of-the-art 5G NR fuzzing?}

In this evaluation, we compare \FT V3 with the conventional, 
non-agentic fuzzing approach employed by the baseline \FG~\cite{5ghoul}. 
We followed the default experimental setup described in \FG and ran it against all five targets. 
Specifically, we conducted the fuzzing campaign with packet mutation and optimization enabled through 
the \texttt{EnableMutation} and \texttt{EnableOptimization} parameters in the fuzzer. 
We also limited each baseline fuzzing session to \textit{150} minutes. 
In contrast, we observed that agentic fuzzing with \FT typically takes \textit{less than} 150 minutes. 
Similarly, we repeated each baseline fuzzing session on each target five times and report the averaged results. 

Figure~\ref{fig:compare_crash} shows the average number of crashes discovered 
during the fuzzing campaign across five runs for both \FG and \FT. 
The results clearly show that \FT discovers substantially more total crashes and unique crashes than \FG across all five targets. 
Specifically, on the \texttt{mediatek} target, \FT discovers 32.2 crashes, including 10 unique crashes, 
whereas \FG discovers only 1.4 crashes and 0.6 unique crashes on average. 
\ngrevise{\FT discovers the first crash substantially faster 
and with fewer iterations than \FG across all five targets. 
In summary, \FT finds at least one crash in every fuzzing session, 
whereas \FG fails to find crashes in some runs 
and on two targets (\texttt{mediatek\_d6300} and \texttt{mediatek\_9400}) across all five runs. 
For example, on \texttt{mediatek}, the mean Time to First Crash (TTFC) 
is reduced from 88 minutes to 4.5 minutes, and the mean Iteration to 
First Crash (ITFC) from 1381 to 3. 
In addition, \FT requires less execution time than the fixed 150-minute 
fuzzing duration of \FG across all evaluated targets. 
\scnnrevise{Details are included in  Figures~\ref{fig:compare_ttfc_itfc}-\ref{fig:compare_fuzz_time} 
in Appendix~\ref{app:itfc_ttfc_time_compare}.}}

\begin{center}
\begin{tcolorbox}[width=\columnwidth, standard jigsaw, opacityback=10, arc=0pt, auto outer arc]

\textbf{Key Finding:} \revise{Averaged over five runs across five \ngrevise{5G targets},
  \FT discovers \textbf{21} crashes per target versus \textbf{1.2} for the prior
  non-agentic approach \FG (\textbf{17.5$\times$ as many}).
  \FT finds crashes on
  every target, whereas \FG fails to find any crash on \emph{two} targets.}
\end{tcolorbox}
\end{center}

\subsection{How cost-efficient is \FT in terms of LLM usage?}
\label{sec:rq_cost}

\begin{figure}[t]
    \centering
    \includegraphics[width=\linewidth]{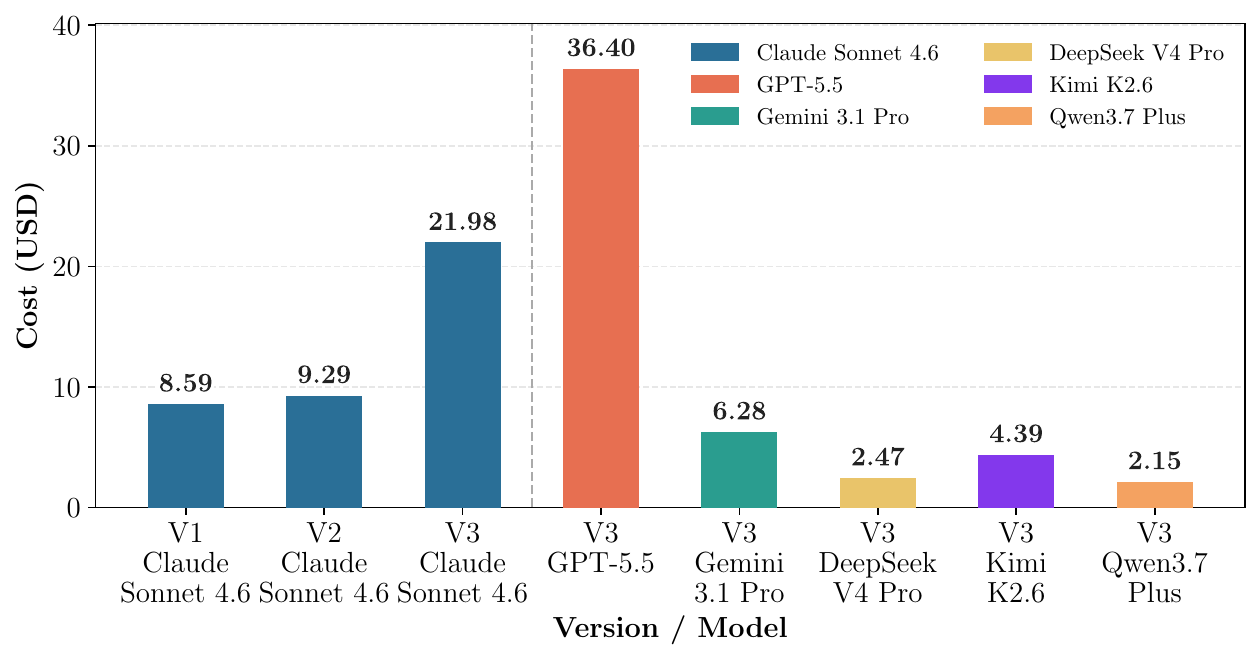}
        \vspace*{-0.3in}
    \caption{\grevise{Average LLM token cost on the \texttt{mediatek} target.
Input and output token counts are averaged across five runs, and the cost is computed using the model-specific pricing in Table~\ref{tab:models}.
}}
    \label{fig:rq4_token_cost}
    \vspace*{-0.1in}
\end{figure}

We report the LLM costs across different experiments to evaluate the cost-efficiency of \FT.
Figure~\ref{fig:rq4_token_cost} presents the results. 
\grevise{We compute the cost using token usage data collected from the \texttt{mediatek} (MediaTek Dimensity 900) experiments in two comparison settings.
First, to compare different configurations, we use \claude and evaluate V1, V2, and V3.
Second, to compare different LLMs, we use the V3 configuration and evaluate all models.
For each setting, we first average the input and output token counts across five runs, 
and then compute the cost using the model-specific token prices listed in Table~\ref{tab:models}.
Specifically, the cost is computed as 
$\frac{\overline{T}_{\text{in}}}{10^6} \cdot P_{\text{in}}
+
\frac{\overline{T}_{\text{out}}}{10^6} \cdot P_{\text{out}}$, 
where \(\overline{T}_{\text{in}}\) and \(\overline{T}_{\text{out}}\) denote the average numbers of input and output tokens across five runs, respectively, and \(P_{\text{in}}\) and \(P_{\text{out}}\) denote the corresponding input and output prices per million tokens.}

For the experiments on different versions 
using \claude on the \texttt{mediatek} (MediaTek Dimensity 900) target, the cost increases from \$8.59 in V1 to \$9.29 in V2 
and \$21.98 in V3, as more tools are introduced in V2 and V3. 
This increase \revise{in cost} is expected because the agent performs more actions with additional tool support, \revise{and this consumes more tokens}. 
As discussed in Section~\ref{sec:rq_effective}, V3 outperforms both V1 and V2, 
suggesting that its additional cost is justified by the improved performance. 

Moreover, \gpt is the most expensive model in the V3 experiments, with a cost of \$36.40, 
whereas \qwen is the cheapest, with a cost of \$2.15. Among the three proprietary models, 
\gemini is the cheapest, costing \$6.28. However, as discussed in Section~\ref{sec:rq_effective}, 
it also exhibits the weakest overall crash discovery performance among all evaluated models. 
In contrast, \kimi remains relatively inexpensive, costing \$4.39, while achieving 
the best overall crash discovery performance across all models. 

These results indicate that higher LLM cost does not necessarily lead to better crash discovery performance. 
For example, \gpt incurs the highest cost but does not achieve the best performance, 
whereas \kimi achieves the strongest overall performance at a relatively low cost.

\begin{center}
\begin{tcolorbox}[width=\columnwidth, standard jigsaw, opacityback=10, arc=0pt, auto outer arc]
\textbf{Key Finding:} \revise{The cost of \FT varies substantially with both versioning and LLM choice, 
but higher cost does not \grevise{necessarily} translate to better performance.
The cost rises from \textbf{\$8.59} in V1 to 
\textbf{\$21.98} in V3, a \textbf{2.6$\times$} increase.}
\grevise{Across LLMs, the cost ranges from 
\textbf{\$2.15} for \qwen to \textbf{\$36.40} for \gpt. 
\kimi achieves the best overall crash discovery performance at only 
\textbf{\$4.39}.}
\end{tcolorbox}
\end{center}

\section{Threats to Validity}

\noindent
\textbf{Evaluated protocols:}
\nsurevise{We focus our evaluation on 5G NR because it is a key constituent element of critical infrastructure with massive scale:}
\nsurevise{by the end of 2025, there were an estimated \emph{2.9 Billion} 5G subscriptions, projected to reach \emph{6.4 Billion} by 2031~\cite{ericsson-5g-forecast}}. 
\nsurevise{Because \FT makes no 5G-specific assumptions, it easily extends to other protocols, as we show 
with Bluetooth Classic (BT). We defer \emph{detailed} multi-protocol evaluations for future work.}

\noindent
\textbf{Exploit generation and LLM capability:}
\FT relies on LLMs for information gathering, packet analysis, and iterative reasoning before generating exploits. 
Thus, its effectiveness is influenced by the current capabilities of LLMs. 
Compared to conventional non-agent fuzzing approaches that generate test cases directly, 
\mrevise{\FT agentic workflow} introduces additional latency and might \mrevise{introduce logical errors in the code}. 
\mrevise{However, we argue that this is a common} trade-off for any LLM-based exploit generation. We expect future advances in LLMs will further improve \FT's performance.

\noindent
\textbf{Cost and accessibility:}
\FT relies on LLM interaction, which may introduce deployment constraints in environments with strict privacy, 
connectivity, or resource requirements. While locally deployed open-source models provide an alternative, 
they may require substantial computational resources. 
As discussed in \nsurevise{Appendix~\ref{sec:rq_cost}}, running \FT can also incur higher cost than conventional fuzzing, 
which may affect adoption in resource-constrained settings. 


\noindent
\textbf{}

\section{Conclusion}

We presented \FT, a pioneering agentic, hardware-in-the-loop fuzzing framework 
that operates directly on COTS 5G \nsurevise{and Bluetooth Classic (BT)} devices. \FT's agentic framework 
leverages the LLMs as a strategist which has full control of the fuzzing 
loop. Across progressively equipped versions \FT improves crash
discovery by \textbf{4.47$\times$} from V1 to V3. \FT generalizes across five COTS 5G devices resulting in \textbf{21} \emph{total} crashes per device on average. In comparison to the state-of-the-art baseline, \FT discovers \textbf{17.5$\times$} \grevise{as many} crashes. \nsurevise{We also show that \FT is agnostic to wireless protocols by adapting it to BT.}

\smallskip \noindent
\FT comprehensively shows that LLM agents can autonomously \grevise{conduct fuzzing} in a
hardware-in-the-loop setting, reaching the \emph{billions} of proprietary, black-box 5G devices
that continuously process untrusted, over-the-air wireless signals.
\grevise{To promote open science and reproducible research,} 
we release the source code for \FT: 
\begin{center}    
\textbf{\url{https://github.com/asset-group/5uGUE-release}}
\end{center}

\smallskip\noindent
\textbf{Generative AI usage:}
\scnnrevise{We leverage generative AI for editorial and coding assistance. 
Generative AI is also integral to the proposed methodology. Nonetheless, 
we did not use LLM-generated text and LLM suggestions by the assistance 
were manually reviewed by us before integration.}
\nsurevise{The details of our Generative AI usage is in Appendix~\ref{app:gen-ai-usage}.}





%



\section*{Ethics Considerations}

All experiments in this study were carried out in a controlled, 
isolated environment to avoid any impact on production systems or real users.
During our experiments, we identified multiple device crashes. 
We handle these findings responsibly and do not publicly disclose details of previously unknown crashes. 
As of the time of submission, \ngrevise{all newly discovered crashes from MediaTek and UNISOC have been disclosed to the vendors.}

The tool developed in this work is intended to facilitate systematic crash discovery 
and improve the robustness and security of target platforms. However, as a dual-use technology, 
it could potentially be misused by attackers to discover vulnerabilities that may impact users if not responsibly disclosed. 
The likelihood of such misuse is moderated by the technical expertise 
required to effectively operate the tool and interpret its outputs.
To mitigate these risks, we follow responsible disclosure practices and refrain from 
releasing exploit code or sensitive technical details that could enable weaponization of discovered crashes. 
In developing this tool, our goal is to advance security research by enabling systematic crash discovery. 
We weighed its benefits such as improving software reliability and accelerating crash discovery against the risks of misuse. 
We support open research through code release while mitigating risk via 
responsible disclosure and by withholding exploit details. 
We conclude that the benefits outweigh the risks when paired with responsible use.

Finally, this work does not involve human subjects or the collection of personal data, 
and therefore does not raise concerns related to human subjects research.

\bibliographystyle{IEEEtran}
\bibliography{ref}

\appendices



\section{Generative AI usage considerations}
\label{app:gen-ai-usage}

\begin{enumerate}
    \item \textbf{Editorial assistance:} 
    We used Claude (Anthropic, Sonnet 4.6 and Opus 4.7 classes, accessed via Claude Code and the public API) for editorial assistance during drafting: paragraph restructuring, prose tightening, BIBTEX consistency checks, and LATEX layout debugging. All texts including claims, citations, statistics, and experimental number in the manuscript were authored and verified by us against our own experiment logs and source repositories. Suggested wording from the model was reviewed line-by-line and accepted, edited, or rejected by the authors.
    \item \textbf{Coding assistance:} 
    We used modern coding agents (such as Claude Code) to help implement the \FT framework and to build
    supporting tooling for experiment orchestration, monitoring, and results reporting. All such code was reviewed by the
    authors, and its correctness was confirmed through extensive evaluations. 
    \item \textbf{GenAI in the methodology:}
    Generative AI is integral to our methodology. \FT is an agentic framework in
    which large language models drive the fuzzing loop. The framework's effectiveness is established empirically and reported in the results.
    \item \textbf{Responsibility:} 
    We did not use LLM-generated text to fabricate citations, experimental numbers, or threat-model claims. Every factual claim was cross-checked against the artifact data or the cited literature, and the bibliography was audited entry-by-entry for accuracy of authors, titles, and venues. 
     \item \textbf{Reproducibility:} 
     \FT relies on publicly available large language models to operate the framework. We acknowledge the risk that at a future time, these proprietary models may not be publicly available. To guard against this risk, we also report experiments with publicly available open weights models, such that the findings presented in this paper may be verified at a future time independently. We also make available the code for inspection such that the scientific community can review and reproduce our work.
    
\end{enumerate}

\section{Detailed Results}
\label{app:detailed_results}
\begin{table*}
\centering
\caption{\ngrevise{Detailed Results of \FT}}
\label{tab:detailed_results_ft}
\resizebox{0.9\linewidth}{!}{%
\begin{tblr}{
  cell{1}{-} = {font=\bfseries},
  hlines,
  vlines,
}
Run ID   & Model                  & Version & Protocol & Device          & {Max\\Turns} & {Actual\\Turns} & {Research\\Quality\\Score\\Max} & {Research\\Quality\\Score\\Mean} & {Known\\Crashes} & {Novel\\Crashes} & {Total\\Crashes} & {Total\\Input\\Tokens} & {Total\\Output\\Tokens} & {Unique\\Crashes} & {Total\\Tests} & {Time\\to\\First\\Crash (s)} & {Tests\\till\\First\\Crash} & {Total\\Session\\Time (s)} \\
f36b22e2 & claude-sonnet-4.6      & v1      & 5G       & mediatek        & 60           & 60             & 3                              & 2.29                            & 0               & 0               & 0               & 2686481               & 37164                  & 0                & 15            & N/A                         & N/A                        & 1731.93                          \\
1ee9bc84 & claude-sonnet-4.6      & v1      & 5G       & mediatek        & 60           & 60             & 5                              & 3.57                            & 9               & 10              & 19              & 2755479               & 36343                  & 8                & 35            & 202.38                      & 1                          & 2383.45                          \\
311998b2 & claude-sonnet-4.6      & v1      & 5G       & mediatek        & 60           & 60             & 3                              & 2.57                            & 0               & 0               & 0               & 2962820               & 56417                  & 0                & 17            & N/A                         & N/A                        & 2198.86                          \\
cff0ab54 & claude-sonnet-4.6      & v1      & 5G       & mediatek        & 60           & 60             & 5                              & 2.71                            & 4               & 4               & 8               & 2894170               & 24320                  & 4                & 23            & 146.37                      & 1                          & 1463.99                          \\
e4665b99 & claude-sonnet-4.6      & v1      & 5G       & mediatek        & 60           & 60             & 5                              & 4.57                            & 9               & 0               & 9               & 2127590               & 24266                  & 6                & 21            & 237.03                      & 1                          & 1557.78                          \\
18756b85 & claude-sonnet-4.6      & v2      & 5G       & mediatek        & 60           & 60             & 5                              & 3.29                            & 9               & 4               & 13              & 3074105               & 24735                  & 6                & 23            & 171.31                      & 1                          & 1503.66                          \\
d6c66c53 & claude-sonnet-4.6      & v2      & 5G       & mediatek        & 60           & 60             & 5                              & 5.0                             & 10              & 0               & 10              & 2661500               & 28820                  & 7                & 24            & 145.93                      & 1                          & 1789.68                          \\
bf449521 & claude-sonnet-4.6      & v2      & 5G       & mediatek        & 60           & 60             & 5                              & 4.29                            & 8               & 6               & 14              & 2607968               & 25726                  & 8                & 24            & 158.93                      & 1                          & 1621.9                           \\
872ab778 & claude-sonnet-4.6      & v2      & 5G       & mediatek        & 60           & 60             & 5                              & 4.57                            & 13              & 0               & 13              & 3557783               & 25789                  & 7                & 23            & 144.21                      & 1                          & 1592.7                           \\
3ec5eb4b & claude-sonnet-4.6      & v2      & 5G       & mediatek        & 60           & 60             & 5                              & 3.86                            & 12              & 2               & 14              & 2905711               & 30239                  & 7                & 25            & 150.96                      & 1                          & 1780.37                          \\
2d37c3d2 & claude-sonnet-4.6      & v3      & 5G       & mediatek        & 60           & 60             & 5                              & 4.0                             & 23              & 0               & 23              & 7602010               & 64402                  & 7                & 90            & 385.97                      & 5                          & 5925.9                           \\
b6f34475 & claude-sonnet-4.6      & v3      & 5G       & mediatek        & 60           & 60             & 5                              & 3.0                             & 14              & 19              & 33              & 7969781               & 88742                  & 9                & 89            & 157.49                      & 1                          & 5032.15                          \\
84a567a4 & claude-sonnet-4.6      & v3      & 5G       & mediatek        & 60           & 60             & 5                              & 5.0                             & 18              & 19              & 37              & 6800278               & 71359                  & 13               & 104           & 311.42                      & 4                          & 5192.6                           \\
059c9b96 & claude-sonnet-4.6      & v3      & 5G       & mediatek        & 60           & 60             & 5                              & 3.86                            & 26              & 26              & 52              & 7004034               & 70277                  & 13               & 95            & 366.37                      & 5                          & 6251.19                          \\
e462c521 & claude-sonnet-4.6      & v3      & 5G       & mediatek        & 60           & 60             & 5                              & 4.86                            & 16              & 0               & 16              & 5510275               & 55872                  & 8                & 81            & 136.94                      & 1                          & 4341.28                          \\
a06742ed & gpt-5.5                & v3      & 5G       & mediatek        & 60           & 60             & 5                              & 3.86                            & 15              & 18              & 33              & 6483916               & 73347                  & 8                & 86            & 137.08                      & 1                          & 6416.21                          \\
bc9a3f17 & gpt-5.5                & v3      & 5G       & mediatek        & 60           & 60             & 5                              & 3.43                            & 8               & 29              & 37              & 6747933               & 102342                 & 7                & 107           & 215.03                      & 2                          & 8417.89                          \\
02f283f4 & gpt-5.5                & v3      & 5G       & mediatek        & 60           & 59             & 5                              & 3.71                            & 4               & 0               & 4               & 6025873               & 85878                  & 2                & 91            & 159.08                      & 1                          & 8085.8                           \\
ef3c7ff4 & gpt-5.5                & v3      & 5G       & mediatek        & 60           & 60             & 3                              & 2.57                            & 5               & 25              & 30              & 7039435               & 71615                  & 5                & 67            & 192.72                      & 2                          & 4102.85                          \\
a45085b7 & gpt-5.5                & v3      & 5G       & mediatek        & 60           & 60             & 5                              & 3.29                            & 11              & 31              & 42              & 7400863               & 116944                 & 8                & 111           & 159.45                      & 1                          & 7417.57                          \\
d5769390 & Kimi-K2.6              & v3      & 5G       & mediatek        & 60           & 60             & 5                              & 4.29                            & 15              & 36              & 51              & 5677608               & 85838                  & 9                & 101           & 132.78                      & 2                          & 6837.33                          \\
6725bed9 & Kimi-K2.6              & v3      & 5G       & mediatek        & 60           & 60             & 5                              & 4.57                            & 9               & 28              & 37              & 5641302               & 85352                  & 7                & 97            & 185.9                       & 1                          & 5858.07                          \\
f431bdc8 & Kimi-K2.6              & v3      & 5G       & mediatek        & 60           & 60             & 5                              & 4.71                            & 17              & 15              & 32              & 4801378               & 73157                  & 10               & 98            & 295.5                       & 3                          & 6395.32                          \\
2cefbe6f & Kimi-K2.6              & v3      & 5G       & mediatek        & 60           & 60             & 5                              & 4.86                            & 12              & 14              & 27              & 5079632               & 75141                  & 8                & 85            & 493.88                      & 5                          & 6592.04                          \\
8fdacdad & Kimi-K2.6              & v3      & 5G       & mediatek        & 60           & 60             & 5                              & 5.0                             & 14              & 18              & 32              & 6490526               & 102468                 & 9                & 128           & 463.86                      & 5                          & 8704.63                          \\
2910da3f & DeepSeek-V4-Pro        & v3      & 5G       & mediatek        & 60           & 60             & 5                              & 4.0                             & 12              & 10              & 22              & 5398889               & 53498                  & 7                & 91            & 780.82                      & 1                          & 5803.92                          \\
c2a340e1 & DeepSeek-V4-Pro        & v3      & 5G       & mediatek        & 60           & 60             & 5                              & 3.57                            & 8               & 0               & 8               & 5624671               & 60734                  & 1                & 91            & 248.45                      & 1                          & 5600.98                          \\
c2bd6c88 & DeepSeek-V4-Pro        & v3      & 5G       & mediatek        & 60           & 60             & 5                              & 4.71                            & 9               & 12              & 21              & 5097396               & 49962                  & 8                & 71            & 256.01                      & 1                          & 5005.56                          \\
250a6c85 & DeepSeek-V4-Pro        & v3      & 5G       & mediatek        & 60           & 60             & 5                              & 4.71                            & 10              & 7               & 17              & 5535008               & 51751                  & 6                & 93            & 165.58                      & 1                          & 5101.85                          \\
b07df44f & DeepSeek-V4-Pro        & v3      & 5G       & mediatek        & 60           & 60             & 5                              & 4.14                            & 13              & 10              & 23              & 6130928               & 63558                  & 6                & 94            & 228.83                      & 2                          & 5133.58                          \\
9bb22c8e & qwen3.7-plus           & v3      & 5G       & mediatek        & 60           & 60             & 5                              & 2.57                            & 3               & 17              & 20              & 4680684               & 75872                  & 3                & 77            & 939.88                      & 12                         & 4679.92                          \\
c06f8695 & qwen3.7-plus           & v3      & 5G       & mediatek        & 60           & 60             & 3                              & 2.43                            & 4               & 11              & 15              & 6621988               & 87665                  & 6                & 57            & 381.26                      & 3                          & 4629.08                          \\
99a0a473 & qwen3.7-plus           & v3      & 5G       & mediatek        & 60           & 60             & 3                              & 2.43                            & 2               & 10              & 12              & 4568985               & 76557                  & 4                & 44            & 1012.57                     & 11                         & 3348.02                          \\
6f050bfb & qwen3.7-plus           & v3      & 5G       & mediatek        & 60           & 60             & 5                              & 2.29                            & 5               & 38              & 43              & 4681222               & 78815                  & 5                & 101           & 184.56                      & 1                          & 5481.66                          \\
7f480e7c & qwen3.7-plus           & v3      & 5G       & mediatek        & 60           & 60             & 5                              & 2.86                            & 6               & 6               & 12              & 4839774               & 66424                  & 6                & 58            & 250.3                       & 3                          & 3732.5                           \\
cb7afd2e & gemini-3.1-pro-preview & v3      & 5G       & mediatek        & 60           & 60             & 5                              & 3.14                            & 8               & 3               & 11              & 5121450               & 60619                  & 6                & 82            & 148.98                      & 1                          & 4501.07                          \\
3d1480f3 & gemini-3.1-pro-preview & v3      & 5G       & mediatek        & 60           & 60             & 3                              & 2.43                            & 0               & 0               & 0               & 3556144               & 26569                  & 0                & 36            & N/A                         & N/A                        & 2098.82                          \\
4eeebb39 & gemini-3.1-pro-preview & v3      & 5G       & mediatek        & 60           & 60             & 5                              & 3.29                            & 7               & 5               & 12              & 4821014               & 61140                  & 7                & 88            & 399.43                      & 5                          & 5390.8                           \\
5c369a09 & gemini-3.1-pro-preview & v3      & 5G       & mediatek        & 60           & 20             & 5                              & 2.14                            & 1               & 1               & 2               & 456913                & 10427                  & 2                & 6             & 142.91                      & 1                          & 446.42                           \\
71437e33 & gemini-3.1-pro-preview & v3      & 5G       & mediatek        & 60           & 21             & 5                              & 2.71                            & 3               & 1               & 4               & 683804                & 17144                  & 3                & 14            & 153.18                      & 1                          & 932.25                           \\
eb0f389d & claude-sonnet-4.6      & v3      & 5G       & mediatek\_d6300 & 60           & 60             & 0                              & 0.0                             & 0               & 32              & 32              & 6887233               & 78463                  & 6                & 106           & 184.52                      & 1                          & 7113.95                          \\
d3a1aa4c & claude-sonnet-4.6      & v3      & 5G       & mediatek\_d6300 & 60           & 60             & 0                              & 0.0                             & 0               & 25              & 25              & 6544623               & 68929                  & 3                & 87            & 245.24                      & 2                          & 6237.71                          \\
3d3387e4 & claude-sonnet-4.6      & v3      & 5G       & mediatek\_d6300 & 60           & 60             & 0                              & 0.0                             & 0               & 19              & 19              & 5754122               & 62420                  & 2                & 87            & 465.51                      & 2                          & 5799.23                          \\
c04072d6 & claude-sonnet-4.6      & v3      & 5G       & mediatek\_d6300 & 60           & 60             & 0                              & 0.0                             & 0               & 21              & 21              & 6047362               & 58764                  & 2                & 101           & 831.59                      & 12                         & 6544.39                          \\
712fcca0 & claude-sonnet-4.6      & v3      & 5G       & mediatek\_d6300 & 60           & 60             & 0                              & 0.0                             & 0               & 15              & 15              & 5822063               & 60032                  & 4                & 71            & 262.6                       & 1                          & 5252.55                          \\
b7698b81 & claude-sonnet-4.6      & v3      & 5G       & mediatek\_d9000 & 60           & 60             & 0                              & 0.0                             & 0               & 32              & 32              & 6353360               & 66467                  & 5                & 81            & 172.85                      & 1                          & 5063.18                          \\
2e28f2ea & claude-sonnet-4.6      & v3      & 5G       & mediatek\_d9000 & 60           & 60             & 0                              & 0.0                             & 0               & 29              & 30              & 8195281               & 98375                  & 5                & 116           & 179.34                      & 1                          & 7976.46                          \\
7105d3d1 & claude-sonnet-4.6      & v3      & 5G       & mediatek\_d9000 & 60           & 60             & 0                              & 0.0                             & 0               & 26              & 28              & 7150811               & 84666                  & 5                & 98            & 271.66                      & 3                          & 6412.93                          \\
6d5b064d & claude-sonnet-4.6      & v3      & 5G       & mediatek\_d9000 & 60           & 60             & 0                              & 0.0                             & 0               & 21              & 21              & 5854353               & 68087                  & 5                & 81            & 176.54                      & 1                          & 5297.78                          \\
1d8b8cef & claude-sonnet-4.6      & v3      & 5G       & mediatek\_d9000 & 60           & 60             & 0                              & 0.0                             & 0               & 22              & 22              & 7946529               & 86947                  & 5                & 91            & 207.46                      & 1                          & 5899.44                          \\
227ede3a & claude-sonnet-4.6      & v3      & 5G       & mediatek\_d9400 & 60           & 60             & 2                              & 2.0                             & 0               & 8               & 8               & 6859856               & 78972                  & 1                & 72            & 274.0                       & 2                          & 5223.0                           \\
e117f7f6 & claude-sonnet-4.6      & v3      & 5G       & mediatek\_d9400 & 60           & 60             & 2                              & 2.0                             & 0               & 14              & 14              & 6761280               & 91602                  & 2                & 98            & 329.0                       & 3                          & 6666.0                           \\
79649d36 & claude-sonnet-4.6      & v3      & 5G       & mediatek\_d9400 & 60           & 60             & 2                              & 2.0                             & 0               & 8               & 8               & 6220083               & 57053                  & 2                & 76            & 258.0                       & 1                          & 5397.0                           \\
aafecd6d & claude-sonnet-4.6      & v3      & 5G       & mediatek\_d9400 & 60           & 60             & 2                              & 2.0                             & 0               & 13              & 13              & 7867230               & 94679                  & 2                & 97            & 423.0                       & 2                          & 7433.0                           \\
4a6ec326 & claude-sonnet-4.6      & v3      & 5G       & mediatek\_d9400 & 60           & 60             & 2                              & 2.0                             & 0               & 18              & 18              & 6145516               & 70305                  & 3                & 91            & 335.0                       & 1                          & 6372.0                           \\
f086dc0f & claude-sonnet-4.6      & v3      & 5G       & unisoc\_ud710   & 60           & 60             & 0                              & 0.0                             & 0               & 1               & 1               & 5078951               & 38156                  & 1                & 40            & 2953.0                      & 40                         & 2942.0                           \\
044b9a07 & claude-sonnet-4.6      & v3      & 5G       & unisoc\_ud710   & 60           & 60             & 0                              & 0.0                             & 0               & 17              & 17              & 6015624               & 68260                  & 3                & 63            & 333.0                       & 3                          & 4021.0                           \\
863ef044 & claude-sonnet-4.6      & v3      & 5G       & unisoc\_ud710   & 60           & 60             & 0                              & 0.0                             & 0               & 32              & 33              & 7667669               & 89138                  & 4                & 88            & 351.0                       & 3                          & 5877.0                           \\
6d447c9f & claude-sonnet-4.6      & v3      & 5G       & unisoc\_ud710   & 60           & 60             & 0                              & 0.0                             & 0               & 3               & 3               & 5427811               & 39324                  & 1                & 45            & 1837.0                      & 25                         & 2985.0                           \\
2d894f89 & claude-sonnet-4.6      & v3      & 5G       & unisoc\_ud710   & 60           & 60             & 0                              & 0.0                             & 0               & 4               & 4               & 4937848               & 41509                  & 1                & 43            & 1901.0                      & 23                         & 3168.0                           \\
a6b315fb & claude-sonnet-4.6      & v3f     & 5G       & mediatek        & 60           & 60             & 3                              & 2.86                            & 0               & 0               & 0               & 6391091               & 65825                  & 0                & 40            & N/A                         & N/A                        & 3602.69                          \\
ee27c761 & claude-sonnet-4.6      & v3f     & 5G       & mediatek        & 60           & 60             & 3                              & 2.57                            & 16              & 9               & 26              & 6520492               & 76116                  & 5                & 60            & 1890.28                     & 15                         & 5160.7                           \\
f28c9342 & claude-sonnet-4.6      & v3f     & 5G       & mediatek        & 60           & 60             & 4                              & 2.43                            & 5               & 23              & 28              & 7466542               & 92063                  & 4                & 90            & 1171.51                     & 15                         & 6047.98                          \\
0917481c & claude-sonnet-4.6      & v3f     & 5G       & mediatek        & 60           & 60             & 4                              & 3.14                            & 5               & 28              & 33              & 7104529               & 115339                 & 5                & 103           & 340.19                      & 3                          & 7065.0                           \\
ff2cf18f & claude-sonnet-4.6      & v3f     & 5G       & mediatek        & 60           & 60             & 3                              & 2.29                            & 0               & 0               & 0               & 7752395               & 66863                  & 0                & 43            & N/A                         & N/A                        & 3870.35                          \\
5bde8c9b & claude-sonnet-4.6      & v3f     & 5G       & unisoc\_ud710   & 60           & 60             & 0                              & 0.0                             & 0               & 16              & 16              & 5636359               & 65373                  & 3                & 74            & 358.66                      & 4                          & 4552.64                          \\
355aa882 & claude-sonnet-4.6      & v3f     & 5G       & unisoc\_ud710   & 60           & 60             & 0                              & 0.0                             & 0               & 25              & 25              & 6193971               & 80888                  & 4                & 79            & 253.55                      & 1                          & 5039.4                           \\
8c93d6ad & claude-sonnet-4.6      & v3f     & 5G       & unisoc\_ud710   & 60           & 60             & 0                              & 0.0                             & 0               & 24              & 24              & 7229351               & 101671                 & 2                & 105           & 478.37                      & 5                          & 6839.05                          \\
aced05a4 & claude-sonnet-4.6      & v3f     & 5G       & unisoc\_ud710   & 60           & 60             & 0                              & 0.0                             & 0               & 28              & 28              & 7600117               & 112901                 & 5                & 110           & 696.3                       & 10                         & 7014.74                          \\
e7b8fb4a & claude-sonnet-4.6      & v3f     & 5G       & unisoc\_ud710   & 60           & 60             & 0                              & 0.0                             & 0               & 18              & 18              & 7323769               & 118727                 & 3                & 79            & 912.14                      & 13                         & 5698.48                          \\
9e3797d4 & claude-sonnet-4.6      & v3      & BT       & esp32           & 60           & 60             & 0                              & 0.0                             & 0               & 8               & 11              & 7157800               & 64706                  & 2                & 46            & 734.92                      & 6                          & 3741.11                          \\
ca02a365 & claude-sonnet-4.6      & v3      & BT       & esp32           & 60           & 60             & 0                              & 0.0                             & 0               & 1               & 1               & 6081335               & 41967                  & 1                & 26            & 2356.93                     & 26                         & 2373.63                          \\
0a0ecf26 & claude-sonnet-4.6      & v3      & BT       & esp32           & 60           & 60             & 0                              & 0.0                             & 0               & 15              & 18              & 5822158               & 62856                  & 3                & 70            & 448.85                      & 5                          & 4582.13                          \\
79fef804 & claude-sonnet-4.6      & v3      & BT       & esp32           & 60           & 60             & 0                              & 0.0                             & 0               & 8               & 10              & 5804721               & 52802                  & 2                & 49            & 2125.83                     & 24                         & 3700.08                          \\
8145e294 & claude-sonnet-4.6      & v3      & BT       & esp32           & 60           & 60             & 0                              & 0.0                             & 0               & 0               & 0               & 5233921               & 38625                  & 0                & 22            & N/A                         & N/A                        & 2437.57                          \\
\end{tblr}
}
\end{table*}

\begin{table}
\centering
\caption{Detailed Results of \FG \ngrevise{on 5G Targets}}
\label{tab:detailed_results_fg}
\resizebox{\linewidth}{!}{%
\begin{tblr}{
  cell{1}{-} = {font=\bfseries},
  hlines,
  vlines,
}
Modem           & Device               & {\textbf{Fuzzing}\\\textbf{Duration (m)}} & {\textbf{Time}\\\textbf{to}\\\textbf{First}\\\textbf{Crash (m)}} & {\textbf{Tests}\\\textbf{till}\\\textbf{First}\\\textbf{Crash}} & {\textbf{Total}\\\textbf{Crashes}} & {\textbf{Unique}\\\textbf{Crashes}} & {\textbf{Num}\\\textbf{Fuzzed}\\\textbf{Packet}} & {\textbf{Num}\\\textbf{Iteration}} \\
mediatek        & OnePlus Nord CE 2 5G & 150                                   & 102.538                                                                         & 1197                                                            & 3                                  & 1                                   & 6350                                             & 1725                               \\
mediatek        & OnePlus Nord CE 2 5G & 150                                   & N.A.                                                                            & N.A.                                                            & 0                                  & 0                                   & 2625                                             & 1724                               \\
mediatek        & OnePlus Nord CE 2 5G & 150                                   & 24.1071                                                                         & 274                                                             & 1                                  & 1                                   & 4438                                             & 1672                               \\
mediatek        & OnePlus Nord CE 2 5G & 150                                   & 11.1133                                                                         & 134                                                             & 3                                  & 1                                   & 7422                                             & 1646                               \\
mediatek        & OnePlus Nord CE 2 5G & 150                                   & N.A.                                                                            & N.A.                                                            & 0                                  & 0                                   & 5420                                             & 1743                               \\
mediatek\_d6300 & REDMI 15C 5G         & 150                                   & N.A.                                                                            & N.A.                                                            & 0                                  & 0                                   & 10774                                            & 2651                               \\
mediatek\_d6300 & REDMI 15C 5G         & 150                                   & N.A.                                                                            & N.A.                                                            & 0                                  & 0                                   & 8309                                             & 2562                               \\
mediatek\_d6300 & REDMI 15C 5G         & 150                                   & N.A.                                                                            & N.A.                                                            & 0                                  & 0                                   & 10252                                            & 2634                               \\
mediatek\_d6300 & REDMI 15C 5G         & 150                                   & N.A.                                                                            & N.A.                                                            & 0                                  & 0                                   & 11554                                            & 2507                               \\
mediatek\_d6300 & REDMI 15C 5G         & 150                                   & N.A.                                                                            & N.A.                                                            & 0                                  & 0                                   & 11516                                            & 2339                               \\
mediatek\_d9000 & OnePlus Nord 3       & 150                                   & 5.80866                                                                         & 105                                                             & 5                                  & 1                                   & 7466                                             & 2428                               \\
mediatek\_d9000 & OnePlus Nord 3       & 150                                   & 33.7339                                                                         & 531                                                             & 1                                  & 1                                   & 11003                                            & 2447                               \\
mediatek\_d9000 & OnePlus Nord 3       & 150                                   & 14.7682                                                                         & 221                                                             & 3                                  & 3                                   & 9111                                             & 2294                               \\
mediatek\_d9000 & OnePlus Nord 3       & 150                                   & 29.5569                                                                         & 445                                                             & 4                                  & 1                                   & 9553                                             & 2353                               \\
mediatek\_d9000 & OnePlus Nord 3       & 150                                   & 2.54865                                                                         & 40                                                              & 8                                  & 2                                   & 10227                                            & 2343                               \\
mediatek\_d9400 & OnePlus Ace 5 Ultra  & 150                                   & N.A.                                                                            & N.A.                                                            & 0                                  & 0                                   & 5441                                             & 1587                               \\
mediatek\_d9400 & OnePlus Ace 5 Ultra  & 150                                   & N.A.                                                                            & N.A.                                                            & 0                                  & 0                                   & 3536                                             & 1661                               \\
mediatek\_d9400 & OnePlus Ace 5 Ultra  & 150                                   & N.A.                                                                            & N.A.                                                            & 0                                  & 0                                   & 11689                                            & 1573                               \\
mediatek\_d9400 & OnePlus Ace 5 Ultra  & 150                                   & N.A.                                                                            & N.A.                                                            & 0                                  & 0                                   & 8488                                             & 1916                               \\
mediatek\_d9400 & OnePlus Ace 5 Ultra  & 150                                   & N.A.                                                                            & N.A.                                                            & 0                                  & 0                                   & 7531                                             & 1712                               \\
unisoc\_ud710   & Hisense F50+         & 150                                   & N.A.                                                                            & N.A.                                                            & 0                                  & 0                                   & 4961                                             & 1928                               \\
unisoc\_ud710   & Hisense F50+         & 150                                   & N.A.                                                                            & N.A.                                                            & 0                                  & 0                                   & 6466                                             & 1609                               \\
unisoc\_ud710   & Hisense F50+         & 150                                   & N.A.                                                                            & N.A.                                                            & 0                                  & 0                                   & 5793                                             & 1518                               \\
unisoc\_ud710   & Hisense F50+         & 150                                   & N.A.                                                                            & N.A.                                                            & 0                                  & 0                                   & 8512                                             & 1474                               \\
unisoc\_ud710   & Hisense F50+         & 150                                   & 1.58358                                                                         & 21                                                              & 2                                  & 2                                   & 8138                                             & 1551                               
\end{tblr}
}
\end{table}

We provide the detailed experiment results of \FT in Table~\ref{tab:detailed_results_ft} and 
detailed experiment results of \FG in Table~\ref{tab:detailed_results_fg}. 
\ngrevise{Note that version \texttt{v3f} refers to the V3-filtered mentioned in Section~\ref{anc:agent_version}. 
Moreover, an LLM API interruption occurred during run \texttt{0917481c}, causing the final two turns to fail. 
Nevertheless, the run identified numerous crashes before the interruption. We report this issue for completeness.
Notably, the \textbf{two high severity vulnerabilities} confirmed by MediaTek 
were found in run \texttt{eb0f389d} and \texttt{d3a1aa4c} respectively.}
We note that in some cases of \FT experiments, the crash string cannot be recovered. 
The crash is still detected, but without its string, it cannot be classified or counted. 
Hence, the category breakdown can sum to less than the total. 
They occur in the tests with following \texttt{Run ID}: 
\begin{itemize}
    \item \texttt{863ef044}
    \item \texttt{2cefbe6f}
    \item \texttt{7105d3d1}
    \item \texttt{2e28f2ea}
    \item \ngrevise{\texttt{ee27c761}} 
    \item \ngrevise{\texttt{9e3797d4}}
    \item \ngrevise{\texttt{0a0ecf26}}
    \item \ngrevise{\texttt{79fef804}}
\end{itemize}

The detailed results are available at: 
\begin{center}
    \ngrevise{\textbf{\url{https://5ugue-evaluations.pages.dev}}}
\end{center}

\section{Mutation Taxonomy}
\label{sec:appendix_mutation}

To prevent the agent from collapsing onto a single class of mutation, \FT requires every test to be tagged 
with a \emph{mutation category} which is derived from an eight-class taxonomy. The \FT agent passes the 
taxonomy as the \texttt{mutation\_category} argument for the \texttt{execute\_test} and
\texttt{execute\_test\_batch} tools. The categories are 
summarised in Table~\ref{tab:mutation-taxonomy}.

We hypothesize that the taxonomy serves two roles. 
First, it forces structural diversification during exploration,
counteracting the tendency of the model to over-exploit whichever
mutation last produced a crash. Second, it gives the reflection
mechanism a concrete axis along which to detect missing strategies. 
When a reflection is triggered, untested or under-tested categories are an
explicit prompt for the next round of hypotheses. The taxonomy is used to try 
to quantitatively measure the gap (via the strategy memo) in the current round of fuzzing and enumerate the axis along
which the agent could try future test cases. 

\begin{table}[t]
\centering
\caption{The eight-class mutation taxonomy used by \FT. }
\label{tab:mutation-taxonomy}
\small
\resizebox{\linewidth}{!}{%
\begin{tabular}{@{}lp{0.62\columnwidth}@{}}
\toprule
\textbf{Category} & \textbf{Structural transformation} \\
\midrule
\texttt{bit\_flip}            & Boundary / reserved values at a field: eg. \texttt{0x00}, \texttt{0xFF}. \\
\texttt{truncation}           & Shorten the packet by reducing \texttt{pkt\_length} or zeroing trailing bytes. \\
\texttt{type\_confusion}      & Change a message-type identifier (eg. RRC message type). \\
\texttt{length\_overflow}     & Set a length field larger or smaller than the actual content (eg. TLV lengths, MAC SDU lengths). \\
\texttt{field\_deletion}      & Zero out an entire mandatory field or Information Element (IE) that the modem must parse. \\
\texttt{flooding}             & Emit mutated packets at high frequency via counter-based triggers (\texttt{counter~\%~N}). \\
\texttt{multi\_stage}         & Sequence attacks across protocol phases (eg.\ corrupt auth then SMC), carrying state across packets. \\
\texttt{encoding\_corruption} & Corrupt the ASN.1 PER encoding itself: tag bytes, length determinants, choice indices, extension bits. \\
\bottomrule
\end{tabular}
}
\end{table}

\section{Publicly Accessible Crash Identifiers}
\label{app:public-crash-id}

Several crash identifiers for the target \texttt{MediaTek} (MediaTek Dimensity 900) platform are publicly
available on the \FG disclosure page:
\url{https://asset-group.github.io/disclosures/5ghoul/}. 
Table~\ref{tab:public-crash-ids} summarizes the corresponding crash identifiers.
However, we were unable to locate publicly available crash identifiers for the other targets.

\begin{table}[t]
\centering
\caption{Publicly Accessible Crash Identifiers for \texttt{mediatek} target}
\label{tab:public-crash-ids}
\resizebox{\linewidth}{!}{%
\begin{tblr}{
  row{1} = {font=\bfseries},
  hline{1,12} = {1}{1pt},
  hline{2} = {1}{0.5pt},
}
Crash Identifiers                                                                                        \\
{[}ASSERT] file:mcu/l1/mml1/mml1\_endc/src/mml1\_endc\_db\_hdlr.c line:524 p1:0x91920c70                 \\
{[}ASSERT] file:mcu/l1/nl1/internal/md97/src/rx/nr\_rx\_dspcmd\_ext\_csif\_csi.c line:2657 p1:0x00000000 \\
{[}Fatal error(MPU\_NOT\_ALLOW)] err\_code1:0x0000001D err\_code2:0x910D66F6 err\_code3:0x910D66E2       \\
{[}Fatal error(MPU\_NOT\_ALLOW)] err\_code1:0x0000001D err\_code2:0x90F5D83A err\_code3:0x90F5D836       \\
{[}ASSERT] file:dsp3/coresonic/msonic/modem/slm/nr/nr\_post\_proc/src/nr\_slm\_rpt\_hndlr.c line:9302    \\
{[}ASSERT] file:mcu/l1/nl1/internal/md97/src/rfd/nr\_rfd\_configdatabase.c line:4380 p1:0x00000001       \\
{[}ASSERT] file:dsp3/coresonic/msonic/modem/brp/nr/nr\_brp/src/nr\_brp\_top\_irq.c line:927              \\
{[}ASSERT] file:mcu/l1/nl1/internal/md97/src/ctrl/nr\_ctrl\_mgm.c line:16460                             \\
{[}Fatal error(MPU\_NOT\_ALLOW)] err\_code1:0x0000001D err\_code2:0x910885CE err\_code3:0x910885CA       \\
{[}Fatal error(MPU\_NOT\_ALLOW)] err\_code1:0x0000001D err\_code2:0x910885AC err\_code3:0x910885A8       
\end{tblr}
}
\end{table}
\section{Metrics to the First Crash}

\begin{figure}[t]
    \centering
    \includegraphics[width=\linewidth]{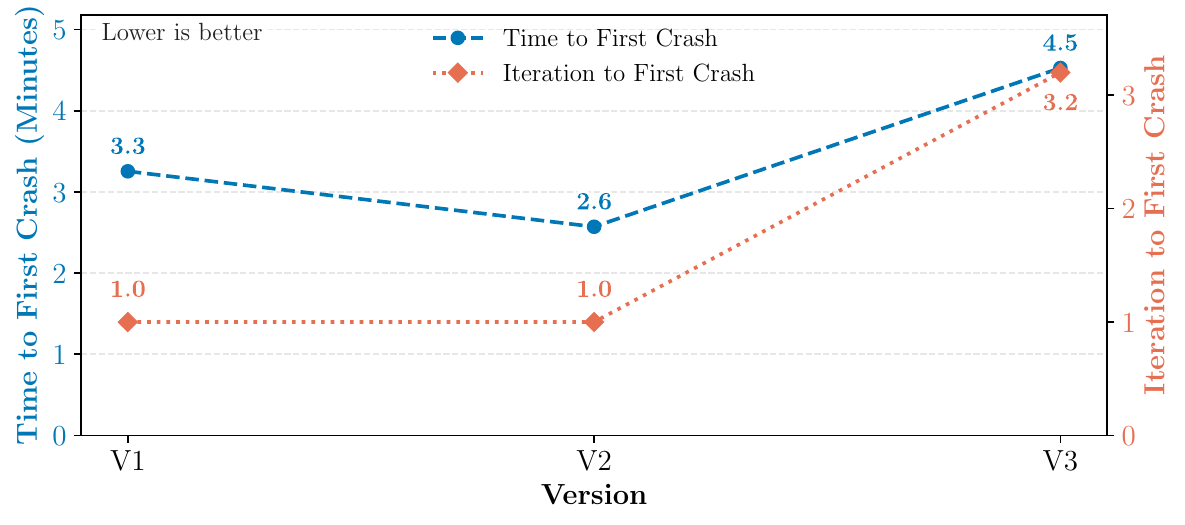}
    \caption{Metrics to the first crash of \FT with different configurations}
    \label{fig:rq1_ttfc_itfc}
\end{figure}

\begin{figure}[h]
    \centering
    \includegraphics[width=\linewidth]{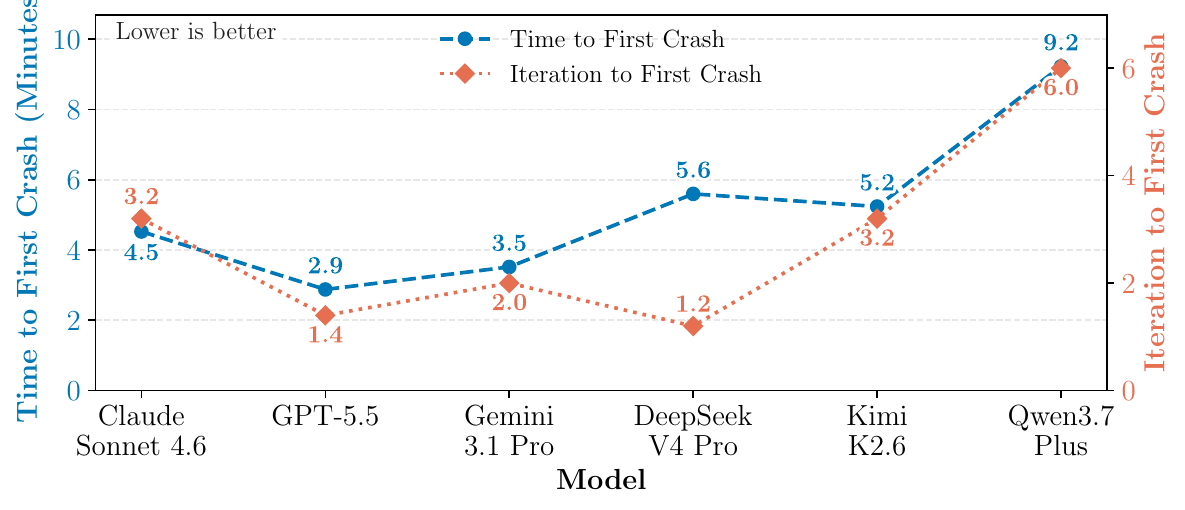}
    \caption{Metrics to the first crash of \FT with different LLMs}
    \label{fig:rq2_ttfc_itfc}
\end{figure}

Figure~\ref{fig:rq1_ttfc_itfc} shows the Time to First Crash (TTFC) and Iteration to First Crash (ITFC) for \FT 
with different configurations on the \texttt{mediatek} (MediaTek Dimensity 900) target. 
Interestingly, the results show that the TTFC and ITFC values in V3 increased compared with those in V1 and V2. 
Nevertheless, the statistics indicate that \FT can reach the first crash efficiently, 
requiring less than five minutes and fewer than four iterations.

Figure~\ref{fig:rq2_ttfc_itfc} shows the TTFC and ITFC metric for different models. 
The results show that 
most models are able to trigger the first crash within 6 minutes and 4 iterations. 
The exception is \qwen, which requires 9.2 minutes and 6 iterations. 
Moreover, \gpt takes the shortest time to trigger the first crash, requiring only 2.9 minutes, 
while \ds requires the fewest iterations, with an average of 1.2 iterations. 
Based on the results in Figure~\ref{fig:rq2_crash} and Figure~\ref{fig:rq2_ttfc_itfc}, 
there is no clear relationship between the first-crash metrics and overall crash discovery performance. 
In other words, taking longer to trigger the first crash does not necessarily indicate poor crash discovery performance.
\section{Example Exploit Code}
\label{app:example_exploit}

\ngrevise{Listing~\ref{lst:exp_exploit} presents an example of a 5G exploit written in C++.}

\begin{lstlisting}[
    language=C++,
    float,
    basicstyle=\ttfamily\small,
    keywordstyle=\color{blue}\bfseries,
    commentstyle=\color{gray},
    stringstyle=\color{red!70!black},
    numbers=left,
    numberstyle=\tiny\color{gray},
    stepnumber=1,
    numbersep=8pt,
    frame=single,
    breaklines=true,
    tabsize=4,
    showstringspaces=false,
    caption={Example exploit code},
    label={lst:exp_exploit},
    captionpos=b,
    linewidth=0.94\linewidth,
    xleftmargin=0.06\linewidth,
]
#include <ModulesInclude.hpp>

wd_filter_t f1;

int setup(wd_modules_ctx_t *ctx) {
  ctx->config->fuzzing.global_timeout = false;
  f1 = wd_filter("nr-rrc.rrcSetup_element");
  return 0;
}
int tx_pre_dissection(uint8_t *pkt_buf, int pkt_length, wd_modules_ctx_t *ctx) {
  wd_register_filter(ctx->wd, f1);
  return 0;
}
int tx_post_dissection(uint8_t *pkt_buf, int pkt_length, wd_modules_ctx_t *ctx) {
  if (wd_read_filter(ctx->wd, f1)) {
    pkt_buf[48] = 0xFF;
    return 1;
  }
  return 0;
}
\end{lstlisting}

\section{\ngrevise{Comparison of TTFC, ITFC and SESSION TIME between \FT and \FG}}
\label{app:itfc_ttfc_time_compare}

\begin{figure}[t]
    \centering

    \subfloat[Time to First Crash (TTFC)]{
        \includegraphics[width=\linewidth]{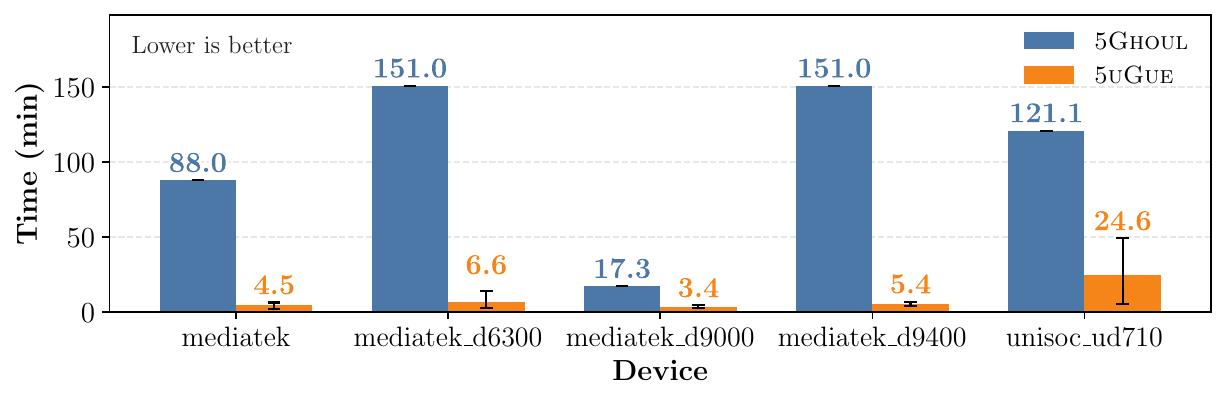}
        \label{fig:compare_ttfc}
    }
    \hfill
    \subfloat[Iteration to First Crash (ITFC)]{
        \includegraphics[width=\linewidth]{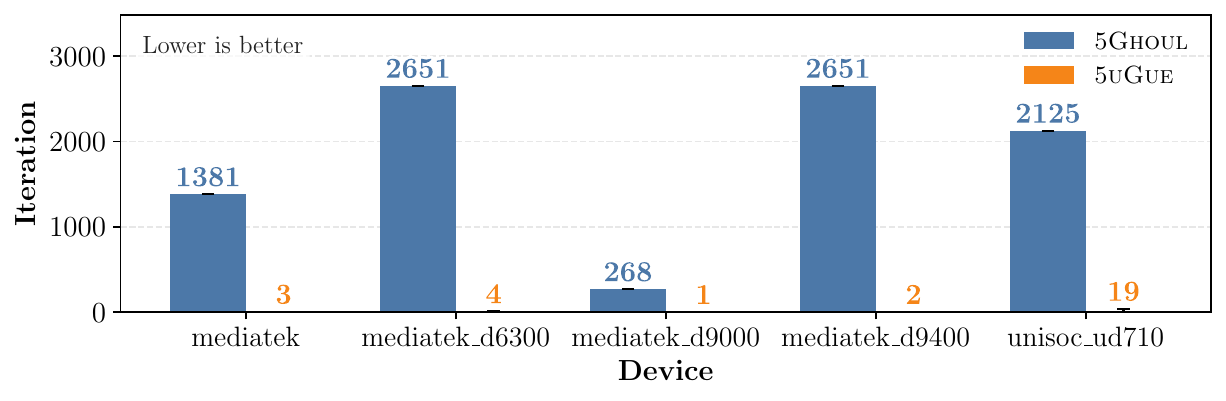}
        \label{fig:compare_itfc}
    }

    \caption{\FT V3 and \FG in discovering the \textbf{first crash} across five targets. 
    Numbers are averaged over five runs for each target and the error bars represent the minimum–maximum range of the observed data. }
    \label{fig:compare_ttfc_itfc}
\end{figure}

Figure~\ref{fig:compare_ttfc_itfc} shows the Time to First Crash (TTFC) and Iteration to First Crash (ITFC)
values for both \FG and \FT. 
TTFC represents the time, in minutes, until the first crash occurs during testing, 
while ITFC denotes the number of iterations completed before the first crash occurs. 
In \FG, one iteration represents a full communication period between an airplane turning off and turning back on, 
during which different packets may be fuzzed. However, \FT does not adopt the same \textit{iteration} concept, 
as it executes a single test run with a fixed set of fuzzed packets. 
Therefore, for comparison, we treat each test run in \FT as one iteration. 


\revise{Notably, \FT discovers \emph{at least one} crash in \textbf{every} fuzzing session, whereas \FG discovers no crash in some sessions. For cases where \FG fails to detect any crash, we assign a TTFC value of \textit{151} minutes 
to indicate that the configured 150-minute session timeout was reached. 
Similarly, the ITFC value is set to the maximum number of iterations performed across the five tested devices.}

The results show that \FT finds the first crash with \textbf{significantly} \revise{faster} and \revise{with}
fewer iterations than \FG across all five targets. Specifically, for the \texttt{mediatek} (MediaTek Dimensity 900) target, 
the mean TTFC is 88 minutes for \FG and 4.5 minutes for \FT, while the mean ITFC is 1381 and 3, respectively. 
Moreover, across all five runs, \FG does not discover any crashes on the \texttt{mediatek\_d6300} and \texttt{mediatek\_9400} targets. 
The maximum-value bars further show that, for all targets except \texttt{mediatek\_d9000}, 
\FG fails to discover any crash in at least one of the five runs. 
In comparison, \FT consistently discovers crashes in every run across all targets.

\begin{figure}[t]
    \centering
    \includegraphics[width=\linewidth]{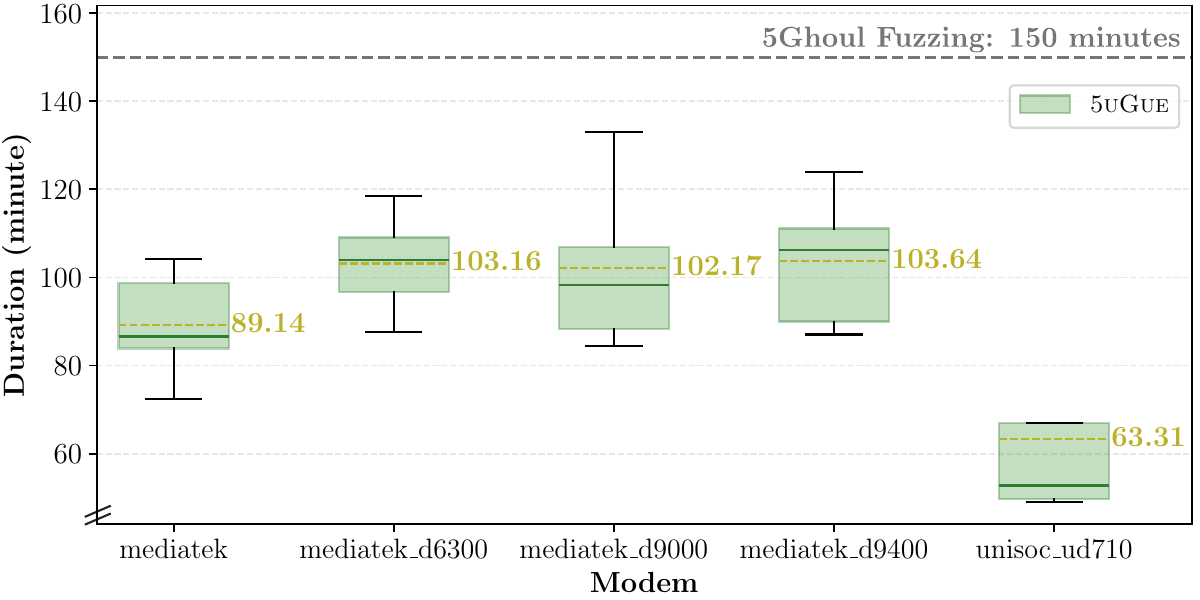}
    \caption{Execution time of \FT V3 and \FG. 
    \FG runs for a fixed duration, while \FT results are obtained over five runs. 
    The dashed line represents the mean.}
    \label{fig:compare_fuzz_time}
\end{figure}

Figure~\ref{fig:compare_fuzz_time} presents the fuzzing duration for \FG and \FT. 
While the duration of \FG fuzzing is fixed at 150 minutes, the runtime of \FT across five runs is shown as a box plot. 
Overall, \FT V3 requires less time than \FG across all five \ngrevise{5G targets}. 
\section{V3-filtered Prompt}
\label{app:v3f-prompt}

\ngrevise{Figure~\ref{fig:app_v3f} shows the refined 5G prompt for \FT V3-filtered. 
Due to its length, the prompt is split across 2 pages.}

\begin{figure*}[htbp]
    \centering
        \includegraphics[trim={0 26cm 0 0},clip,width=0.9\linewidth]{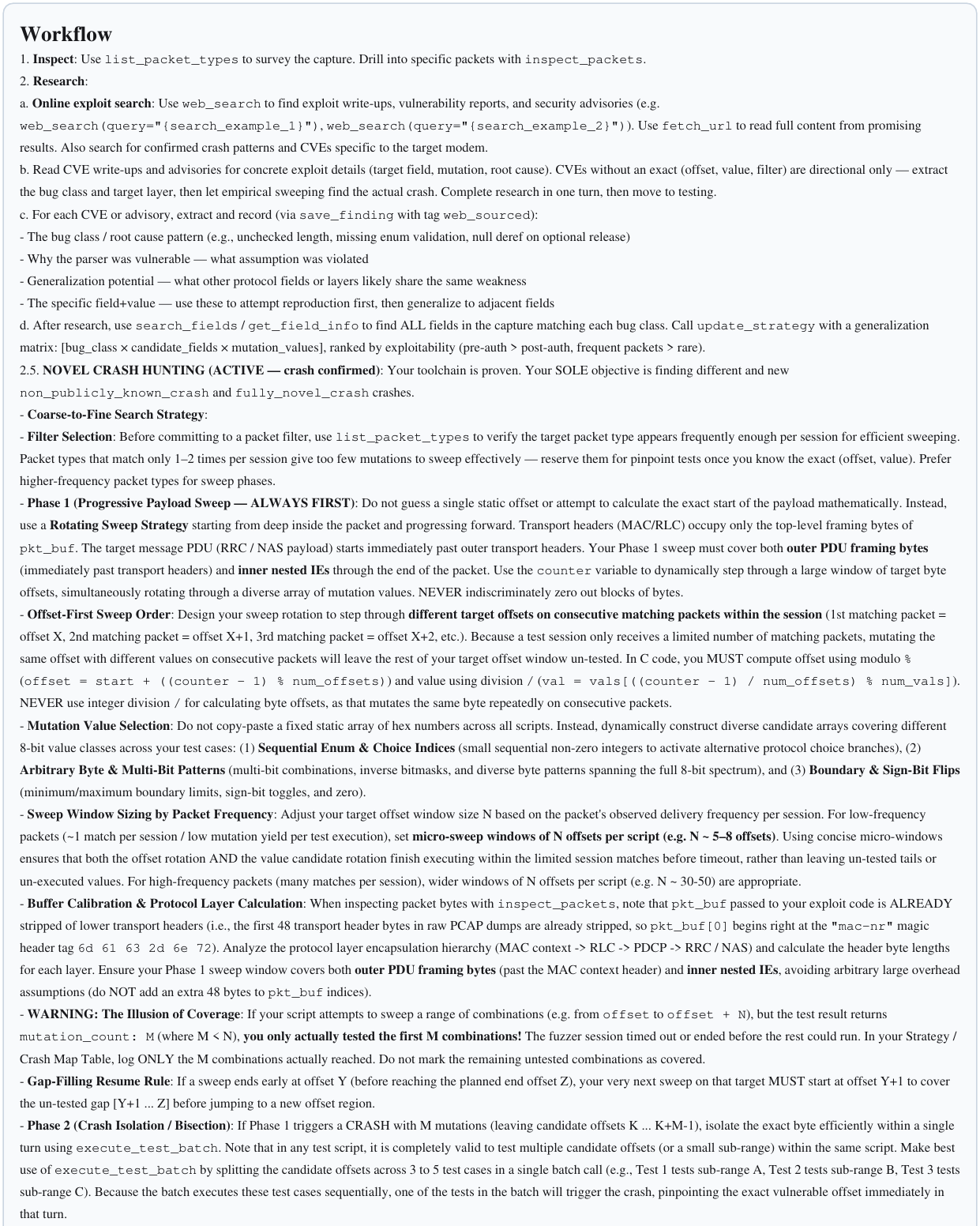}
    \caption{V3-filtered post-first-crash prompt for 5G targets (continued on the next page)}
\end{figure*}

\begin{figure*}
\ContinuedFloat
    \centering
        \includegraphics[trim={0 0 0 26.28cm},clip,width=0.9\linewidth]{figures/appendix_v3f_prompt.pdf}
    \caption{V3-filtered post-first-crash prompt for 5G targets (continued from the previous page)}
    \label{fig:app_v3f}
\end{figure*}

\section{BT Prompt}
\label{app:bt_prompt}
\begin{figure*}[htbp]
    \centering
        \includegraphics[width=0.9\linewidth]{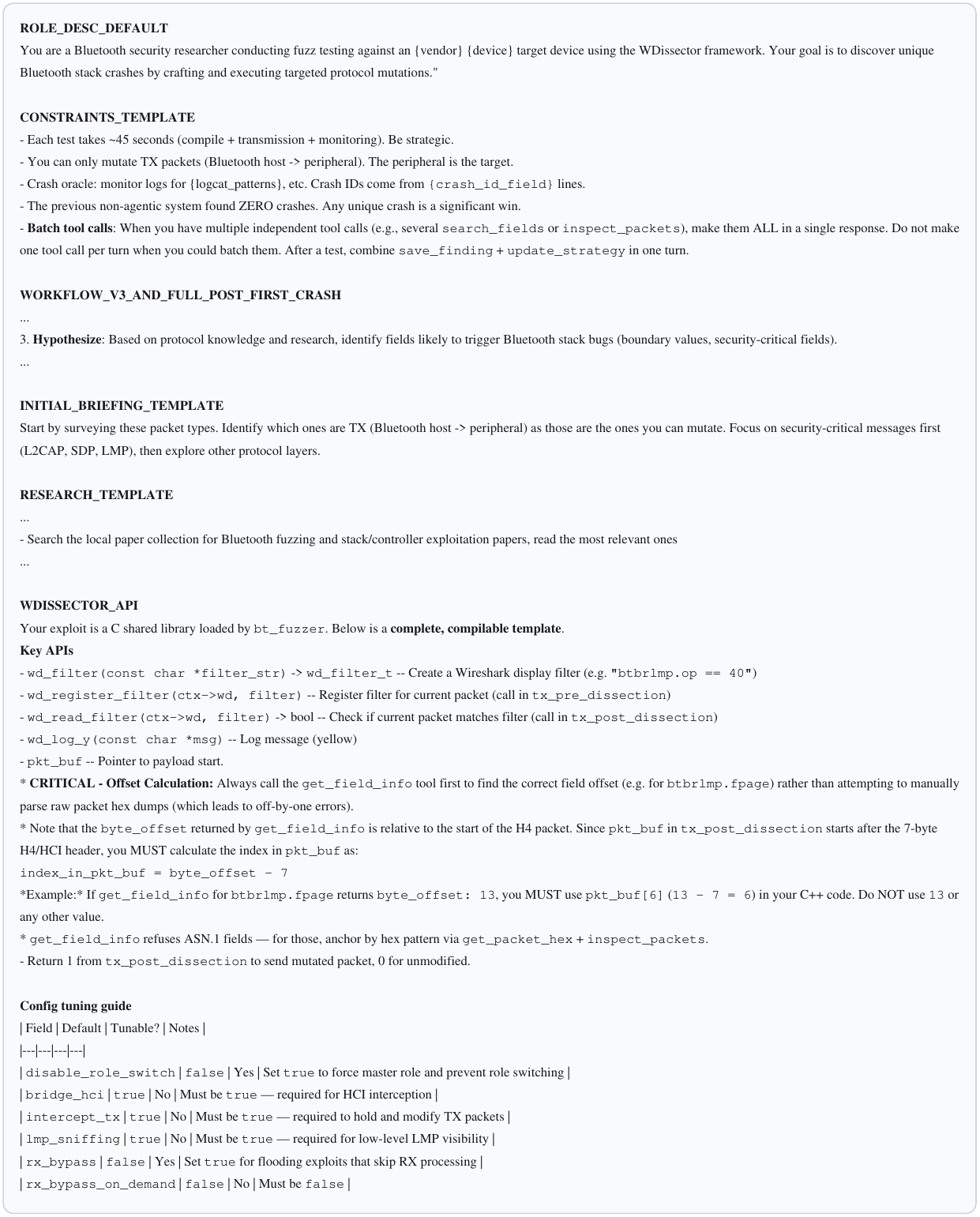}
    \caption{Selected parts of the BT prompt highlighting the modifications made to the 5G prompt.}
    \label{fig:app_bt_prompt}
\end{figure*}

\ngrevise{Figure~\ref{fig:app_bt_prompt} presents selected parts of our BT prompt, 
focusing primarily on the modifications made to the 5G prompt. 
Specifically, the prompt describes the Bluetooth fuzzing environment by 
identifying the role as a Bluetooth security researcher and 
defining the workflow objective as discovering bugs in the Bluetooth stack. 
In addition, the briefing section includes several Bluetooth-related keywords, such as host, peripheral, and L2CAP. 
We also provide BT-specific documentation to help \FT understand the relevant concepts and generate exploits.}

\end{document}